\documentclass[10pt,aps,showpacs,nofootinbib,prd,aps,epsf,floats,
               amsmath,amssymb,amsfonts,axodraw,floatfix,graphicx,twocolumn]{revtex4-1}
\usepackage{amsmath, amssymb}
\usepackage{multirow}
\usepackage{paralist}
\usepackage{slashed}
\usepackage{hyperref} 
\newcommand{\mathsym}[1]{{}}

\usepackage{graphicx}
\usepackage{amsmath}
\usepackage{amssymb}
\usepackage{mathrsfs}
\newsavebox{\PSLASH}
 \sbox{\PSLASH}{$p$\hspace{-1.8mm}/}
 
\renewcommand{\theequation}{\thesection.\arabic{equation}}
\newcounter{saveeqn}
\newcommand{\add}{\addtocounter{equation}{1}}
\newcommand{\alphaeqn}{\setcounter{saveeqn}{\value{equation}}%
\setcounter{equation}{0}%
\renewcommand{\theequation}{\mbox{\thesection.\arabic{saveeqn}{\alpha{equation}}}}}
\newcommand{\reseteqn}{\setcounter{equation}{\value{saveeqn}}%
\renewcommand{\theequation}{\thesection.\arabic{equation}}}

 \newsavebox{\notrightarrow}
 \sbox{\notrightarrow}{$\to$\hspace{-4mm}/}
 
 \newsavebox{\PARTIALSLASH}
 \sbox{\PARTIALSLASH}{$\partial$\hspace{-1.6mm}/}
 
 \newsavebox{\ASLASH}
 \sbox{\ASLASH}{$A$\hspace{-2.1mm}/}
 
 \newsavebox{\KSLASH}
 \sbox{\KSLASH}{$k$\hspace{-1.8mm}/}
 
 \newsavebox{\LSLASH}
 \sbox{\LSLASH}{$\ell$\hspace{-1.8mm}/}
 
 \newsavebox{\QSLASH}
 \sbox{\QSLASH}{$q$\hspace{-1.8mm}/}
 
 \newsavebox{\DSLASH}
 \sbox{\DSLASH}{$D$\hspace{-2.2mm}/}
 
 \newsavebox{\DbfSLASH}
 \sbox{\DbfSLASH}{${\mathbf D}$\hspace{-2.8mm}/}
 
 \newsavebox{\DELVECRIGHT}
 \sbox{\DELVECRIGHT}{$\stackrel{\rightarrow}{\partial}$}
 
 \newcommand{\blue}{\IfColor{\textCadetBlue}{}}
\newcommand{\black}{\IfColor{\textBlack}{}}
\newcommand{\red}{\IfColor{\textRed}{}}
\newcommand{\green}{\IfColor{\textOliveGreen}{}}
\newcommand{\lil}{\IfColor{\textRedViolet}{}}

\newcommand{\bs}{\boldsymbol}
\newcommand{\MD}{MDCDW}

\makeatother
\usepackage[T1]{fontenc}
\usepackage[latin9]{inputenc}
\usepackage{amsmath}
\usepackage{amssymb}
\usepackage{graphicx}
\usepackage{dcolumn}
\usepackage{verbatim}

\usepackage{orcidlink}

\begin{document}
\title{Magnetic dual chiral density wave phase in rotating cold quark matter}
\author{H. Mortazavi Ghalati}\email{hooman.mortazavi@physics.sharif.ir}
\author{N. Sadooghi\,\orcidlink{0000-0001-5031-9675}~~}\email{Corresponding author: sadooghi@physics.sharif.ir}
\affiliation{Department of Physics, Sharif University of Technology,
P.O. Box 11155-9161, Tehran, Iran}
\begin{abstract}
The effect of rotation on the formation of the magnetic dual chiral density wave (\MD) in a dense and magnetized cold quark matter is studied. This phase is supposed to exist in the extreme conditions prevailing, e.g., in a neutron star. These conditions are, apart from high densities and strong magnetic fields, a relatively large angular velocity. To answer the question of whether the rotation enhances or suppresses the formation of this phase, we first determine the effect of rotation on the energy dispersion relation of a fermionic system in the presence of a constant magnetic field and then focus on the thermodynamic potential of the model at low temperature $T$ and finite chemical potential $\mu$. The thermodynamic potential consists, in particular, of an anomalous part leading to certain topological effects. We show that in comparison with the nonrotating case, a term proportional to the angular velocity appears in this anomalous potential. We then solve the corresponding gap equations to the chiral and spatial modulation condensates, and study the dependence of these dynamical variables on the chemical potential ($\mu$), magnetic field ($eB$), and angular velocity ($\Omega$). It turns out that the interplay between these parameters suppresses the formation of the \MD~phase in relevant regimes for cold neutron stars. This is interpreted as the manifestation of the inverse magnetorotational catalysis, which is also reflected in the phase portraits  $eB$-$\mu$, $eB$-$\Omega R$, and $\mu$-$\Omega R$, explored in this work.
\end{abstract}

\maketitle
\section{Introduction}\label{sec1}
\setcounter{equation}{0}
Exploring the phase diagram of quark matter under extreme conditions is one of the important subjects in nuclear physics. These conditions are, among others, high temperature, large baryon chemical potential, and the presence of uniform electromagnetic fields. Big questions related to these subjects are discussed in \cite{rajagopal2018}. Open problems and possible past, present, and future theoretical and experimental strategies to answer these questions are discussed recently in various reports and reviews \cite{pisarski2022, aarts2023, hot-QCD,present2023}. Apart from standard simulation methods in lattice Quantum Chromodynamics (QCD) \cite{aarts2023}, new computational tools, such as machine learning \cite{machine2023}, are developed and applied to either analyze the experimental data or to overcome the deficiencies of standard methods in working with QCD under extreme conditions.
\par
Neutron stars are the natural playground to study quark matter under extreme conditions. These stars, which are produced by gravitational collapse of very massive stars, are the densest objects in the Universe. Their inner density is several times larger than the nuclear saturation mass density $\rho_{n}\sim 2.5\times 10^{17}$ kg/m$^{3}$ (see e.g. \cite{reed2020}). There have been several attempts to determine the equation of states of neutron stars \cite{wambach2023}. The latter is a necessary input for the Toleman-Oppenheimer-Volkoff equation, which together with the mass continuity equation yields the mass-to-radius ratio of these stars \cite{hatsuda-book}. Using effective models, like the Nambu--Jona-Lasinio (NJL) model, it was found that the matter in the interior of neutron stars is in the color superconductivity (CS) phase \cite{rajagopal2007}. In particular, the three-flavor color-flavor locked (CFL) phase seemed to be the most favored phase \cite{rajagopal1999}. The CFL phase, however, does not pass certain astrophysical tests \cite{ferrer2022}, and is thus ruled out.
\par
Apart from large densities, neutron stars exhibit very strong magnetic fields. The strength of the magnetic fields of magnetars is estimated to be $\sim 10^{18}$ G for nuclear matter and $\sim 10^{20}$ G for quark matter (see \cite{ferrer2022} and references therein).
The effect of constant magnetic fields on CS phases is studied, e.g. in \cite{ferrer-MCFL}. In \cite{sadooghi2010}, it is shown that the two-flavor CS phase is also favorable in a magnetized quark matter at intermediate chemical potentials. By increasing the baryon density from low values to densities a few times higher than the nuclear saturation density, another phase can be formed at low temperatures. This phase, originally introduced in \cite{deryagin1992} and studied in several follow-up papers, e.g.\cite{son2000}, is characterized by quark-hole pairs having a finite total momentum and leading to standing waves. In \cite{tatsumi2005}, in analogy to the static spin density waves, known from condensed matter physics, a density wave is introduced in quark matter at moderate densities which is referred to as the ``dual chiral density wave''. It is represented by a dual standing wave in scalar and pseudoscalar condensates. The effect of uniform magnetic fields on the formation of this phase is studied for the first time in \cite{frolov2010}. The corresponding phase is dubbed the \MD~ phase. It is characterized by two dynamical variables, chiral and spatial modulation condensates, which are determined by solving the corresponding gap equations. It is further shown that because of a certain asymmetry appearing in the energy dispersion relations corresponding to the lowest Landau level (LLL), the thermodynamic potential consists of an anomalous term. As it is argued in \cite{ferrer2015,ferrer2017}, the anomalous term leads to certain topological effects that include, among others, an anomalous nondissipative Hall current and an anomalous electric charge. Moreover, this phase is characterized by the formation of a hybridized propagating mode known as an axion-polariton, which has interesting astrophysical consequences discussed extensively in \cite{ferrer2020}. Recently, in \cite{incera2022}, the phase diagram of the \MD~phase was explored at finite temperatures and in the presence of uniform magnetic fields. It is shown that the \MD~phase is favored at magnetic fields and temperatures compatible with neutron stars. In particular, at intermediate densities, where a remnant mass is formed, the spatial modulation increases. This opens the possibility for this phase to be a favorable candidate for the quark matter in neutron stars.
\par
In addition to high densities and large magnetic fields, neutron stars are also characterized by relatively large angular velocities of about $\Omega_{\text{max}}\sim 10^{3}$ Hz. This leads to a linear velocity $\sim 10^{-2}c-10^{-1}c$, where $c$ is the light velocity.
It is the purpose of this paper, to study the effect of rotation on the formation of the \MD~phase. We assume that the magnetic field and angular velocity are uniform. This is only possible for cold neutron stars which are expected to rotate uniformly \cite{rezzolla-book}. This assumption justifies another assumption concerning the temperature in this paper: Here, in contrast to \cite{incera2022}, we neglect the effect of temperature and its possible interplay with high densities, magnetic fields, and rotations on the formation/suppression of the \MD~phase in a cold quark matter.
\par
As it is argued in \cite{fukushima2015, sadooghi2021}, a certain interplay between the effect of rotation and the magnetic field destroys chiral condensates. An effect which is referred to as ``rotational magnetic inhibition''  \cite{fukushima2015} or ``inverse magnetorotational catalysis'' (IMRC) \cite{sadooghi2021}. For the \MD~phase, the consequence of this effect would be a vanishing of the spatial modulation condensate, as a result of a certain correlation between two condensates in this phase. In this paper, we show that this indeed happens. A fact that may rule out this phase to be favorable for quark matter in rotating, dense, magnetized, and cold neutron stars. We emphasize that the present work is a natural extension of \cite{incera2022}, where 
it is argued that the \MD~phase is ``viable candidate for the matter of neutron stars'' \cite{incera2022}. Particularly in this context, we show that the inclusion of rotation to the condition imposed on the model used in \cite{incera2022} suppresses the formation of the \MD~phase. It is not clear how robust the conclusions presented in \cite{incera2022} as well as those in the present paper, are in a more realistic model of a neutron star, where certain conditions, like $\beta$-equilibrium, isospin symmetry, and charge neutrality have to be imposed \cite{bielich2020}. 
\par
The organization of this paper is as follows: In Sec. \ref{sec2}, we start with the Lagrangian density of a two-flavor NJL model with $U(1)_{\text{L}}\times U(1)_{\text{R}}\times SO(2)\times R^3$  symmetry in the presence of a background magnetic field and introduce the rotation by implementing an angular velocity parallel to the magnetic field in the Hamiltonian of the model.  Then, defining two inhomogeneous condensates, we introduce the mass and spatial modulation condensates. We then solve the energy eigenvalue equation and determine, in particular, the energy eigenvalues and eigenfunctions in a cylindrical coordinate system. The solution to the Dirac equation in a rotating fermionic system with an without boundary conditions, in the absence and/or presence of uniform magnetic fields is studied in \cite{ambrus2014,ambrus2015,chernodub2016,chernodub2017}, apart from \cite{fukushima2015,sadooghi2021}. In the present paper, we neglect the effect of boundary conditions.  In Sec. \ref{sec3}, we determine the thermodynamic (one-loop effective) potential at finite temperature and density, and in the presence of constant magnetic fields. In particular, we focus on the low temperature limit as in \cite{frolov2010} and explore the effect of rotation on the anomalous part of the effective potential. Having the thermodynamic potential at hand, it can be minimized with respect to two condensates. In Sec. \ref{sec4A}, we present our numerical results for the dependence of mass and spatial modulation condensates on the chemical potential $\mu$, magnetic field $eB$ and linear velocity $\Omega R$, with $\Omega$ the angular velocity and $R$ the radius. The results confirm the fact that rotation destroys both condensates as a consequence of the IMRC effect. Moreover, it is shown that for $\mu=0$, the role of $\mu$ is played by a pure rotation in interplay with a background magnetic field. In Sec. \ref{sec4B}, we then explore the $eB$-$\mu$, $eB$-$\Omega R$, and  $\mu$-$\Omega R$ phase diagrams to study the impact of the IMRC effect on various phases appearing in the model. We conclude our results in Sec. \ref{sec5}.

\section{The Model}\label{sec2}
\setcounter{equation}{0}
To study the  \MD~phase in a rotating and dense medium, we start with the Lagrangian density of a two-flavor gauged NJL model,
\begin{eqnarray}\label{N1}
\mathscr{L}= \bar{\psi}[i\slashed{\Pi}- m_0] \psi + G [(\bar{\psi}\psi)^2 + (\bar{\psi}i\gamma^5\boldsymbol{\tau}\psi)^2],\nonumber\\
\end{eqnarray}
where $\psi^{T}=\left(u,d\right)$, $m_0\equiv m_u=m_d$ is the up and down quark bare mass, and $G$ the dimensionful NJL coupling constant. In what follows, we set $m_{0}=0$.
Assuming a rigid rotation about the $z$ direction with the angular velocity $\Omega$, and a magnetic field aligned in the same direction, the operator $\slashed{\Pi}$ in \eqref{N1} reads\footnote{To derive $\slashed{\Pi}$ in \eqref{N2}, we follow exactly the same steps as in \cite{fukushima2015,chernodub2016,chernodub2017,sadooghi2021}.},
\begin{eqnarray}\label{N2}
\slashed{\Pi}=\gamma^0(\partial_t - i\Omega J_z) + \gamma^1\partial_x + \gamma^2 \partial_y + \gamma^3 \partial_z +ieQ\gamma^\mu A_\mu,\nonumber\\
\end{eqnarray}
with $e>0$. Plugging the charge matrix $Q\equiv\mbox{diag}\left(q_u,q_d\right)=\mbox{diag}\left(2/3,-1/3\right)$ and the symmetric gauge $A_{\mu}=\left(0,-\boldsymbol{A}\right)=\left(0,By/2, -Bx/2,0\right)$ into \eqref{N2}, we arrive at
\begin{eqnarray}\label{N3}
\slashed{\Pi}_{f}&=&\gamma^0(\partial_t - i\Omega J_z) + \gamma^1\left(\partial_x+iq_{f}eBy/2\right)\nonumber\\
&& + \gamma^2 \left(\partial_y-iq_{f}eBx/2\right) + \gamma^3 \partial_z,
\end{eqnarray}
where the subscript $f=(u,d)$ denotes the two chosen flavors in this model. In this way, the magnetic field $\boldsymbol{B}=B\hat{\boldsymbol{z}}$ with $B>0$ is also  aligned in the $z$ direction. The Dirac $\gamma$ matrices in \eqref{N2} and \eqref{N3} are defined in the Weyl representation
\begin{eqnarray}\label{N4}
\gamma^{0}=\left(
\begin{array}{cc}
0&1\\
1&0
\end{array}
\right), \quad \boldsymbol{\gamma}=\left(
\begin{array}{cc}
0&\boldsymbol{\tau}\\
-\boldsymbol{\tau}&0
\end{array}
\right),  \quad \gamma^{5}=\left(
\begin{array}{cc}
-1&0\\
0&+1
\end{array}
\right), \nonumber\\
\end{eqnarray}
with the Pauli matrices $\boldsymbol{\tau}=\left(\tau_1, \tau_2, \tau_3\right)$. Moreover, the total angular momentum in the $z$ direction is given by $J_z\equiv L_z+\Sigma_z/2$ with $L_z=-i\left(x\partial_y-y\partial_x\right)$, and $\Sigma_z\equiv \mathbb{I}_{2\times 2}\otimes \tau_{3}$.
\par
As is described in  \cite{ferrer2022}, the \MD~phase at finite baryon density is characterized by two nonlocal condensates
\begin{eqnarray}\label{N5}
\langle\bar{\psi}\psi\rangle=\Delta\cos(q\cdot x), \qquad \langle\bar{\psi}i\gamma^{5}\tau_{3}\psi\rangle=\Delta\sin\left(q\cdot x\right),\nonumber\\
\end{eqnarray}
with $\Delta=\mbox{const}$ and $q\cdot x\equiv q_{\mu}x^{\mu}$. For simplicity, we assume that the modulation vector is aligned along the magnetic field direction $\boldsymbol{z}$, $q^{\mu}=(0,0,0,2b)$ \cite{frolov2010, ferrer2022}. Expanding $\left(\bar{\psi}\psi\right)^2$ and $\left(\bar{\psi}i\gamma^{5}\tau_{3}\psi\right)^2$ in the interacting part of the Lagrangian \eqref{N1} around their mean fields
\eqref{N5}, we arrive first at the semi-bosonized Lagrangian
\begin{eqnarray}\label{N6}
\mathscr{L}_{\text{MF}}=\bar{\psi}[i\slashed{\Pi}-me^{-2i\gamma^{5}\tau_{3} \theta}]\psi-\frac{m^{2}}{4G},
\end{eqnarray}
where the dynamical mass $m\equiv -2G\Delta$, with $\Delta$ given in \eqref{N5} and $\theta\equiv bz$. Performing at this stage a local chiral transformation as in \cite{frolov2010, ferrer2022}
\begin{eqnarray}\label{N7}
\psi\to e^{+i\gamma^{5}\tau_{3}\theta}\psi, \qquad \bar{\psi}\to \bar{\psi}e^{+i\gamma^{5}\tau_{3}\theta},
\end{eqnarray}
we arrive at
\begin{eqnarray}\label{N8}
\mathscr{L}_{\text{MF}}=\bar{\psi}[i\slashed{\Pi}_{f}-m+b\gamma^{5}\gamma^{3}\tau_{3}]\psi-\frac{\left(m\right)^{2}}{4G}.\nonumber\\
\end{eqnarray}
Here, we used $\slashed{\partial}\theta=\gamma^{3}b$. Plugging $\tau_{3}$ into \eqref{N8}, we obtain the corresponding Lagrangian density for each flavor $\mathscr{L}_{f}$,
\begin{eqnarray}\label{N9}
\mathscr{L}_{\text{MF}}=\sum\limits_{f=\{u,d\}}\mathscr{L}_{f},
\end{eqnarray}
with
\begin{eqnarray}\label{N10}
\mathscr{L}_{f}=\bar{\psi}_{f}\left(i\slashed{\Pi}_{f}-m+s_{f}b\gamma^{5}\gamma^{3}\right)\psi_{f}.
\end{eqnarray}
Here, $\slashed{\Pi}_{f}$ is defined in \eqref{N3} and $s_{f}\equiv \mbox{sign}(q_{f})$ is given by $(s_{u},s_{d})=(+1,-1)$ for up and down quarks, respectively. To determine the energy spectrum of this model,  we solve the energy eigenfunction equation,
\begin{eqnarray}\label{N11}
\mathscr{H}_{f}\psi_{f}=E_{f}\psi_{f},
\end{eqnarray}
with the Hamiltonian for each flavor $f$,
\begin{eqnarray}\label{N12}
\mathscr{H}_{f}\equiv i\gamma^{0}\boldsymbol{\gamma}\cdot\boldsymbol{\Pi}_{f}+m\gamma^{0}-s_f b\gamma^{0}\gamma^{5}\gamma^{3}.
\end{eqnarray}
Here, $\boldsymbol{\gamma}\cdot\boldsymbol{\Pi}_{f}$ includes the spatial part of $\Pi_{f}^{\mu}$, defined in \eqref{N3}. Following the method presented in App. \ref{appA}, the corresponding energy eigenvalues to the lowest and higher Landau levels (LLL and HLL) read:
\\
- For LLL ($n=0$), we obtain
\begin{eqnarray}\label{N13}
E_{f}^{n=0}=-\Omega j+b+\epsilon\sqrt{p_{z}^{2}+m^2},
\end{eqnarray}
with $\epsilon=\pm 1$. \\
- For HLL ($n>0$), we obtain
\begin{eqnarray}\label{N14}
E_{f}^{n>0}=-\Omega j+\zeta\bigg[\left(b+\epsilon\sqrt{p_{z}^{2}+m^2}\right)^{2}+2n|q_{f}eB|\bigg]^{1/2},\nonumber\\
\end{eqnarray}
with $\epsilon=\pm 1$, $\zeta=\pm 1$, and $n=1,2,3,\cdots$. Moreover, $j\equiv \ell+1/2$ is the eigenvalue of the $J_{z}$ operator. The latter commutes with $\mathscr{H}_{f}$, and has simultaneous eigenfunctions with this operator.
 Comparing \eqref{N14} with the energy eigenvalues of the same model without rotation $E_{f}^{n,\Omega=0}$ in \cite{ferrer2022}, it turns out that $E_{f}^{n,\Omega=0}-\Omega j=E_{f}^{n}$. In what follows, we determine the thermodynamic potential of this model.
\section{The Thermodynamic Potential}\label{sec3}
\setcounter{equation}{0}
To determine the thermodynamic potential of the \MD~model at finite temperature $T$ and chemical potential $\mu$, let us start with
\begin{eqnarray}\label{E1}
V_{\text{eff}}=-\frac{1}{V}\ln Z,
\end{eqnarray}
where $V$ is the four-dimensional space-time volume, and the partition function $Z$ is given by
\begin{eqnarray}\label{E2}
Z=\int\mathscr{D}\bar{\psi}\mathscr{D}{\psi}\exp\left(-i\int d^{4}x \mathscr{L}_{\text{MF}}\right),
\end{eqnarray}
with the mean-field Lagrangian
\begin{eqnarray}\label{E3}
\mathscr{L}_{\text{MF}}=\frac{m^{2}}{4G}-\bar{\psi}\left(i\slashed{\Pi}-m+b\gamma^{5}\gamma^{3}\tau_{3}+\mu\gamma^{0}\right)\psi,\nonumber\\
\end{eqnarray}
that, comparing to $\mathscr{L}_{\text{MF}}$ from \eqref{N8}, we have introduced $\mu$ and neglect the current mass $m_0$. Using then the definition of the Hamiltonian $\mathscr{H}_{f}$ from \eqref{N12}, $V_{\text{eff}}$ is first given by
\begin{eqnarray*}
V_{\text{eff}}=\frac{m^2}{4G}+V_{eff}^{(1)},
\end{eqnarray*}
with 
\begin{eqnarray}\label{E4}
V_{eff}^{(1)}\equiv -\frac{1}{V}\mbox{Tr}_{\text{sfc}}\left\{\ln[i\partial_{0}-\left(\mathscr{H}_{f}-\mu\right)]\right\},
\end{eqnarray}
where the trace is to be built over the spin (s), flavor (f), and color (c) degrees of freedom. To do this, we consider, as in \cite{fukushima2015}, a cylindrical volume with $L_{z}$, the length of the cylinder, and $R$, its radius. Assuming as in \cite{fukushima2015} that the maximum of $\bs{\psi(\rho, \varphi, z)}$  is inside the cylinder, it is possible to determine an upper bound for the summation over $\ell$ (see the Appendix of \cite{fukushima2015}). This leads to the following phase space for the positively and negatively charged quarks in a rotating medium
\begin{eqnarray}\label{E6}
\left\{\begin{array}{ll}
\frac{L_z}{V}\sum\limits_{n=0}^{\infty}~~\sum\limits_{\ell=-n}^{\mathcal{N}_{u}-n-1}\int\frac{dp_z}{2\pi}&\mbox{for $f=u$},\\
\frac{L_z}{V}\sum\limits_{n=0}^{\infty}~~\sum\limits_{\ell=-\mathcal{N}_{d}+n}^{n-1}\int\frac{dp_z}{2\pi}&\mbox{for $f=d$}.\\
\end{array}
\right.
\end{eqnarray}
Here, $n$ and $\ell$ are the quantum numbers corresponding to Landau levels and rotation, respectively. Moreover, $\mathcal{N}_{f}$, the Landau degeneracy factor for each flavor $f=\{u,d\}$ is defined by
\begin{eqnarray}\label{E7}
\mathcal{N}_{f}\equiv \lfloor{ \frac{|q_{f}eB| S}{2\pi} }\rfloor.
\end{eqnarray}
Here, $S=\pi R^2$. Using the replacement
\begin{eqnarray}\label{E8}
p_{0}\to i\omega_{k}\equiv i\pi T(2k+1),~~\mbox{and}~~\int\frac{dp_0}{2\pi}\to iT\sum_{k=-\infty}^{+\infty},\nonumber\\
\end{eqnarray}
and performing the summation over the Matsubara frequencies $\omega_{k}$ by making use of
\begin{eqnarray}\label{E9}
\frac{1}{\beta}\sum_{k}[\omega_{k}^{2}+(E-\mu)^{2}]=|E-\mu|+\frac{2}{\beta}\ln\left(1+e^{-\beta|E-\mu|}\right), \nonumber\\
\end{eqnarray}
with $\beta\equiv T^{-1}$, we arrive at the thermodynamic potential at finite $T$ and $\mu$,
\begin{eqnarray}\label{E10}
V_{\text{eff}}&=&\frac{m^{2}}{4G}-\frac{N_{c}}{4\pi S}\sum\limits_{f}\sum_{(n),\ell}
\nonumber\\
&&\times\int_{-\infty}^{+\infty} dp_z \bigg\{|E_{f}^{n}-\mu|+\frac{2}{\beta}\ln\left(1+e^{-\beta|E_{f}^{n}-\mu|}\right)\bigg\},\nonumber\\
\end{eqnarray}
where
\begin{eqnarray}\label{E11}
\sum\limits_{(n)}\equiv \sum\limits_{n=0,\epsilon}+
\sum\limits_{n>0,\zeta,\epsilon},
\end{eqnarray}
and $N_{c}$ is the number of colors, and $E_{f}^{n}$ for $n=0$ and $n>0$ are given in \eqref{N13} and \eqref{N14}, respectively.
Similar to \cite{frolov2010}, we separate $V_{\text{eff}}^{(1)}$ into the zero $T$ part, $V_{\text{T=0}}$, the part including $\mu$ and $\Omega$, $V_{\mu,\Omega}$, and the finite $T$ part, $V_{T\neq 0}$,
\begin{eqnarray}\label{E12}
V_{\text{eff}}^{(1)}=V_{\text{T=0}}+V_{\mu,\Omega}+V_{T\neq 0},
\end{eqnarray}
with
\begin{eqnarray}\label{E13}
V_{T=0} &=& -\frac{N_c}{4\pi S} \sum_{f}\sum_{(n),\ell} \int dp_z \, |E_{f}^{n}+\Omega j|,\nonumber\\
V_{\mu,\Omega} &=& -\frac{N_c}{4\pi S}\sum_{f}\sum_{(n),\ell} \int dp_z \, (|E_{f}^{n}-\mu|-|E_{f}^{n}+\Omega j|), \nonumber\\
V_{T\neq 0}&=&-\frac{N_c T}{2\pi S}\sum_{f}\sum_{(n),\ell} \int dp_z \,\ln [1+e^{-\beta|E_{f}^{n}-\mu|}].\nonumber\\
\end{eqnarray}
In what follows, we consider only the $T\to 0$ case. Using
$$
\lim\limits_{T\to 0}T\ln\left(1+e^{-\beta x}\right)=-x\theta(-x),
$$
the $V_{T\neq 0}$ vanishes. We thus focus on $V_{T=0}$ and $V_{\mu\Omega}$ in \eqref{E13}. Similar to the nonrotating case, $V_{T=0}$ has to be appropriately regularized.  According to the definition of $E_{f}^{n}$ from \eqref{N13} and \eqref{N14}, $E_{f}^{n}+\Omega j=E_{f}^{\Omega=0}$ is independent of $j=\ell+1/2$ (or equivalently $\ell$). A summation over $\ell$ is therefore possible, and leads to a factor $\mathcal{N}_{f}$, defined in \eqref{E7}. The zero $T$ part of the potential is thus given by
\begin{eqnarray}\label{E14}
\hspace{-0.5cm}V_{T=0}=-\frac{N_{c}}{4\pi S}\sum\limits_{f}\mathcal{N}_{f}\sum\limits_{(n)}\int dp_z |E_{f}^{n,\Omega=0}|.
\end{eqnarray}
Apart from the factor $\mathcal{N}_{f}/S$, this is the same integral that appears also in \cite{frolov2010,incera2022}. Using the same proper-time regularization as in \cite{frolov2010,incera2022}, it reads
\begin{eqnarray}\label{E15}
V_{T=0}=\frac{N_c}{S}\sum_{f}\mathcal{N}_{f}\sum_{(n)}\int dp_z \int_{\frac{1}{\Lambda^2}}^{\infty}  \frac{ds}{\left(4\pi s\right)^{3/2}}e^{-s \left(E_{f}^{n,\Omega=0}\right)^{2}}.\nonumber\\
\end{eqnarray}
Here, $\Lambda$ is a cutoff regulator. It is possible to perform the summation over Landau levels $n$ and $\zeta$ as well as $\epsilon$. Using
\begin{eqnarray}\label{E16}
1+2\sum_{n=1}^{\infty}e^{-2 A n}=\coth\left(A\right),
\end{eqnarray}
the final expression for $V_{T=0}$, is given by
\begin{eqnarray}\label{E17}
V_{T=0}&=&\frac{N_c}{S}\sum_{f=\{u,d\}}\mathcal{N}_{f}\int dp_z \int_{\frac{1}{\Lambda^2}}^{\infty}  \frac{ds}{\left(4\pi s\right)^{3/2}}\nonumber\\
&&\times \sum_{\epsilon=\pm 1}e^{-s\left(b+\epsilon\sqrt{p_{z}^{2}+m^{2}}\right)^{2}}\coth\left(s|q_{f}eB|\right).\nonumber\\
\end{eqnarray}
Let us now consider $V_{\mu,\Omega}$ from \eqref{N12}.
To regularize it, we use as in \cite{frolov2010}, a cutoff regularization by introducing a Heaviside $\theta$ as a function of the regulator $\Lambda^{\prime}$
\begin{eqnarray}\label{E18}
V_{\mu,\Omega}&=&-\frac{N_c}{4\pi S}\sum_{f,(n),\ell}\int_{-\infty}^{+\infty} dp_z \nonumber\\
&&\times(|E_f^{n}-\mu|-|E_f^{n}+\Omega j|)\theta\left(\Lambda^\prime - |E_f^{n}+\Omega j|\right). \nonumber\\
\end{eqnarray}
\begin{widetext}
To evaluate $V_{\mu,\Omega}$, it is necessary to separate it into two parts, corresponding to LLL and HLLs, $n=0$ and $n>0$,
\begin{eqnarray}\label{E19}
V_{\mu,\Omega}=\mathscr{V}_{\mu,\Omega}^{n=0}+\mathscr{V}_{\mu,\Omega}^{n>0},
\end{eqnarray}
with
\begin{eqnarray}\label{E20}
\mathscr{V}_{\mu,\Omega}^{n=0}&\equiv &-\frac{N_{c}}{4\pi S}\sum_{f,\ell,\epsilon}
\int\limits_{-\infty}^{+\infty} dp_z \theta\left(\Lambda^\prime - |E_f^{n=0,\Omega=0}|\right) \left(|E_f^{n=0}-\mu|-|E_f^{n=0,\Omega=0}|\right),\nonumber\\
\mathscr{V}_{\mu,\Omega}^{n>0}&\equiv &-\frac{N_{c}}{4\pi S}\sum_{f,\ell,\zeta,\epsilon,n>0}\int\limits_{-\infty}^{+\infty} dp_z\theta\left(\Lambda^\prime - |E_f^{n>0,\Omega=0}|\right)
 \left(|E_f^{n>0}-\mu|-|E_f^{n>0,\Omega=0}|\right).
\end{eqnarray}
Let us first consider the LLL contribution to the effective potential, $\mathscr{V}_{\mu,\Omega}^{n=0}$. Following the procedure described in \cite{frolov2010}, we arrive first at
\begin{eqnarray}\label{E21}
\mathscr{V}_{\mu,\Omega}^{n=0}&=&-\frac{N_{c}}{4\pi S}\sum_{f,\ell,\epsilon}\int\limits_{-\infty}^{+\infty} dp_z \left(|E_f^{n=0}-\mu|-|E_f^{n=0,\Omega=0}|\right)-\frac{N_{c}b}{\pi S}\sum_{f,\ell}\left(\Omega j+\mu\right).
\end{eqnarray}
For numerical purposes, it is necessary to sum over $\epsilon$ and perform the integration over $p_{z}$ in the first term of $\mathscr{V}_{\mu,\Omega}^{n=0}$. To do this, we use the method described in \cite{frolov2010}, and arrive after some work at\footnote{In App. \ref{appB1}, we outline the derivation of \eqref{E22}.}
\begin{eqnarray}\label{E22}
\hspace{-0.5cm}\mathscr{V}_{f,\ell}\equiv\sum_{\epsilon}\int\limits_{-\infty}^{+\infty} dp_z \left(|E_f^{n=0}-\mu|-|E_f^{n=0,\Omega=0}|\right)=\left\{
\begin{array}{lcl}
0&\qquad&|a_{2}|<|a_{1}|<|m|,\\
-4I(0,P_{a_1})+4|a_{1}|P_{a_{1}}&\qquad&|a_{2}|<|m|<|a_{1}|,\\
4I\left(P_{a_1},P_{a_{2}}\right)+4|a_{1}|P_{a_1}-4|a_{2}|P_{a_{2}}&\qquad&|m|<|a_{2}|<|a_{1}|,\\
0&\qquad&|a_{1}|<|a_{2}|<|m|,\\
4I\left(0,P_{a_{2}}\right)-4|a_{2}|P_{a_{2}}&\qquad&|a_{1}|<|m|<|a_{2}|,\\
4I\left(P_{a_{1}},P_{a_{2}}\right)+4|a_{1}|P_{a_{1}}-4|a_{2}|P_{a_{2}}&\qquad&|m|<|a_{1}|<|a_{2}|,
\end{array}
\right.\nonumber\\
\end{eqnarray}
\end{widetext}
with $a_{1}\equiv b-\Omega j-\mu$, $a_{2}\equiv b$, $P_{a_{i}}=\sqrt{a_{i}^{2}-m^{2}}, i=1,2$, and $I(\Lambda_{1},\Lambda_{2})$ defined in \eqref{appB5}. The second term in \eqref{E21}, is the well-known anomalous potential, $\mathscr{V}_{\text{anomal}}$ \cite{frolov2010, ferrer2022},
\begin{eqnarray}\label{E23}
\mathscr{V}_{\text{anomal}}&\equiv&
-\frac{N_{c}b}{\pi S}\sum_{f,\ell}\left(\Omega j+\mu\right).
\end{eqnarray}
Having in mind that $j=\ell+1/2$, and performing the summation over $f$ and $\ell$ according to \eqref{E6}, we have
\begin{eqnarray}\label{E24}
2\sum_{f,\ell}(\Omega j+\mu)=\big[2\left(\mathcal{N}_{u}+\mathcal{N}_{d}\right)\mu+\left(\mathcal{N}_{u}^{2}-\mathcal{N}_{d}^{2}\right)\Omega\big].\nonumber\\
\end{eqnarray}
The anomalous potential is thus given by
\begin{eqnarray}\label{E25}
\mathscr{V}_{\text{anomal}}=-
\frac{N_{c}b}{2\pi S}\big[2\left(\mathcal{N}_{u}+\mathcal{N}_{d}\right)\mu+\left(\mathcal{N}_{u}^{2}-\mathcal{N}_{d}^{2}\right)\Omega\big].\nonumber\\
\end{eqnarray}
Let us notice that for $\Omega=0$, the result for $\mathscr{V}_{\text{anomal}}$ coincides with the anomalous part of the effective potential presented in \cite{ferrer2022}. According to the above results, after an appropriate cutoff regularization, the LLL contribution to the effective potential $V_{\mu,\Omega}$ is thus given by
\begin{eqnarray}\label{E26}
\mathscr{V}_{\mu,\Omega}^{n=0}=-\frac{N_{c}}{4\pi S}\sum_{f,\ell}\mathscr{V}_{f,\ell}+\mathscr{V}_{\text{anomal}},
\end{eqnarray}
where $\mathscr{V}_{f,\ell}$ is given in \eqref{E22} and $\mathscr{V}_{\text{anomal}}$ in \eqref{E25}.
\par
Let us now consider the HLL contribution to the effective potential, $\mathscr{V}_{\mu,\Omega}^{n>0}$ from \eqref{E20}. We simplify this expression step by step for our numerical purposes. Using the definition of $E_{f}^{n>0}=-\Omega j+\zeta E_{+}$ with
\begin{eqnarray}\label{E27}
E_{+}\equiv\bigg[\left(b+\epsilon\sqrt{p_{z}^{2}+m^{2}}\right)^{2}+2n|q_{f}eB|\bigg]^{1/2},\nonumber\\
\end{eqnarray}
from \eqref{N14}, the summation over $\zeta$ leads to
\begin{eqnarray}\label{E28}
\lefteqn{\sum_{\zeta=\pm 1}\left(|E_f^{n>0}-\mu|-|E_f^{n>0,\Omega=0}|\right)}\nonumber\\
&=&2\left(|\mu+\Omega j|-E_{+}\right)\theta\left(|\mu+\Omega j|-E_{+}\right)\theta\left(\Lambda^{\prime}-E_{+}\right). \nonumber\\
\end{eqnarray}
Plugging this expression into $\mathscr{V}_{\mu,\Omega}^{n>0}$ from \eqref{E20}, it is given by
\begin{eqnarray}\label{E29}
\mathscr{V}_{\mu,\Omega}^{n>0}=-\frac{N_{c}}{2\pi S}\sum_{f,\epsilon}\sum_{n=1}^{N_{\text{max}}}\int_{-\infty}^{+\infty}dp_{z}\sum_{k=0}^{\mathcal{N}_{f}-1}\mathscr{J}_{f,\epsilon,n,k}(p_{z}),\nonumber\\
\end{eqnarray}
with
\begin{eqnarray}\label{E30}
N_{\text{max}}\equiv\bigg\lfloor\frac{\Lambda^{\prime~2}-\left(b+\epsilon\sqrt{p_{z}^{2}+m^{2}}\right)^{2}}{2|q_{f}eB|}\bigg\rfloor,
\end{eqnarray}
which results from $\theta\left(\Lambda^{\prime}-E_{+}\right)$, and
\begin{eqnarray}\label{E31}
\mathscr{J}_{f,\epsilon,n,k}&\equiv& \left(|\mu_{f}+\Omega k|-E_{+}\right)\theta\left(|\mu_{f}+\Omega k|-E_{+}\right)\nonumber\\
&&\times
\theta\left(\Lambda^{\prime}-E_{+}\right).
\end{eqnarray}
Here, $\mu_{f}$ for $f=\{u,d\}$ are defined by
\begin{eqnarray}\label{E32}
\mu_{u}&\equiv&\mu-\Omega\left(n-\frac{1}{2}\right),\nonumber\\
\mu_{d}&\equiv&\mu+\Omega\left(n-\mathcal{N}_{d}+\frac{1}{2}\right).
\end{eqnarray}
They arise by an appropriate redefinition of the summation over $\ell$ for two different flavors in \eqref{E6}. After performing the summation over $k$ in \eqref{E29}, we arrive finally at
\begin{eqnarray}\label{E33}
\mathscr{V}_{\mu,\Omega}^{n>0}=-\frac{N_{c}}{2\pi S}\sum_{f,\epsilon}\sum_{n=1}^{N_{\text{max}}}\int_{-\infty}^{+\infty}dp_{z}\mathscr{K}_{f,\epsilon,n}(p_{z}),\nonumber\\
\end{eqnarray}
where $\mathscr{K}_{f,\epsilon,n}(p_{z})$ is a lengthy conditional expression, presented in Appendix \ref{appB2}. To summarize, $V_{\text{eff}}^{(1)}$ at zero temperature and finite density is thus given by
$$
V_{\text{eff}}^{(1)}=V_{T=0}+V_{\mu,\Omega}=V_{T=0}+\mathscr{V}_{\mu,\Omega}^{n=0}+\mathscr{V}_{\mu,\Omega}^{n>0},
$$
with $V_{T=0}$ given in \eqref{E17}, $\mathscr{V}_{\mu,\Omega}^{n=0}$ in \eqref{E26}, and $\mathscr{V}_{\mu,\Omega}^{n>0}$ in \eqref{E33}. In what follows, we use this effective potential to study the effect of rotation on a dense and magnetized quark matter in the \MD~phase.
\section{Numerical Results}\label{sec4}
\setcounter{equation}{0}
As it is shown in the previous section, the effective potential of a \MD~medium at zero temperature, finite density, and in the presence of rotation is given by
$
V_{\text{eff}}^{(1)}=V_{T=0}+V_{\mu,\Omega}.
$
The regularized expression for $V_{T=0}$, presented in \eqref{E15}, does not depend on $\mu$ and $\Omega$, but depends through $E_{f}^{n,\Omega=0}$ on $m$ and $b$. These are taken as two dynamical variables in this context. The part of the potential including $\mu$ and $\Omega$ is separated into two parts: The LLL contribution $\mathscr{V}_{\mu,\Omega}^{n=0}$, presented in \eqref{E26} includes, in particular, the anomalous potential $\mathscr{V}_{\text{anomal}}$ from \eqref{E25}. This part, together with the HLL contribution $\mathscr{V}_{\mu,\Omega}^{n>0}$, presented in \eqref{E33} depend explicitly on $m$ and $b$. To study the effect of $\mu, eB$ and $\Omega$ on the formation of these two dynamical variables, we have to determine the global minima of $V_{\text{eff}}$ with respect to $m$ and $b$. To do this, we performed a numerical computation in the regime $\mu\in [0,0.8]$ GeV, $\sqrt{eB}\in [0,0.8]$ GeV, corresponding to $eB\in [0,0.64]$ GeV$^{2}$,  and the linear velocity $\Omega R \in [0,0.1]$. To convert the values of $\Omega$ and $eB$ to Hertz (Hz) and Gauss (G), we use $1$ GeV$=1.52\times 10^{24}$ Hz and
$eB=1$ GeV$^{2}$ corresponding to $B\sim 1.7\times 10^{20}$ G.
\par
The main purpose of the present paper, is to answer the question of whether the \MD~phase survives the extreme conditions of cold neutron stars. In particular, what is the effect of rotation on the formation/suppression of this phase in these extreme conditions? In this regard, let us notice that the maximum radius of a neutron star $R_{\text{max}}\sim 10$ km, its maximum angular velocity $\Omega_{\text{max}}\sim 10^{3}$ Hz \cite{fukushima2015}. These lead to a maximum linear velocity $\left(\Omega R\right)_{\text{max}}\sim 3\times 10^{-2} c$.
As concerns the magnetic field of a neutron star, its maximum value is $B_{\text{max}}\sim 10^{17}$ G \cite{dexheimer2021}, which corresponds to $\sqrt{eB}\sim 0.02$ GeV.
The chosen intervals for $\mu,eB$, and $\Omega R$ are thus quite relevant to study the effect of rotation on the \MD~phase in cold, magnetized and rotating neutron stars. In what follows, we work with dimensionless quantities. To do this, all dimensionful variables are rescaled with $\Lambda=1$ GeV. We work, as in \cite{fukushima2015}, with $R=10^3\Lambda^{-1}$, and set, whenever necessary $G\Lambda^2=6$ \cite{frolov2010}, and $\Lambda^{\prime}=10\Lambda$ \cite{frolov2010}.
\par
In Sec. \ref{sec4A}, we present our numerical results for the $\mu,\sqrt{eB}$, and $\Omega R$ dependencies of $m$ and $b$ for different fixed values of $\Omega R, \sqrt{eB}$, and $\mu$. In Sec. \ref{sec4B}, we then present the corresponding phase portraits to $\sqrt{eB}$-$\mu$, $\sqrt{eB}$-$\Omega R$, and $\mu$-$\Omega R$.
\subsection{The effects of $\boldsymbol{\mu,eB}$ and $\boldsymbol{\Omega}$ on the chiral condensate $\boldsymbol{m}$ and spatial modulation $\boldsymbol{b}$}\label{sec4A}
\begin{figure*}[hbt]
	\includegraphics[width=5.5cm, height=4.5cm]{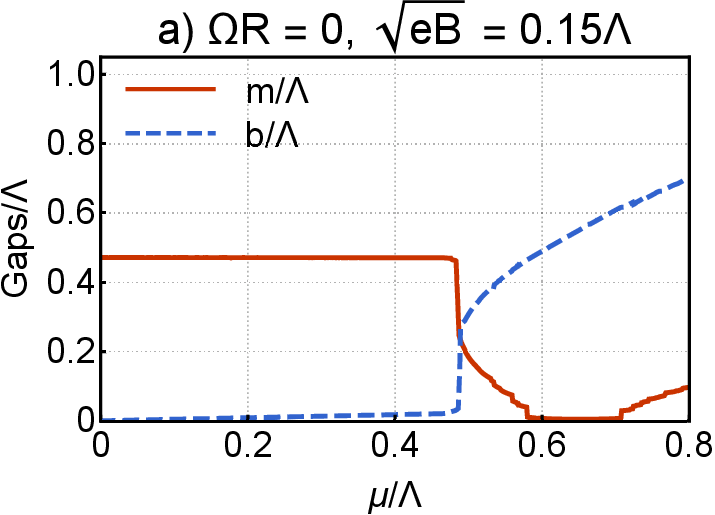}
	\includegraphics[width=5.5cm, height=4.5cm]{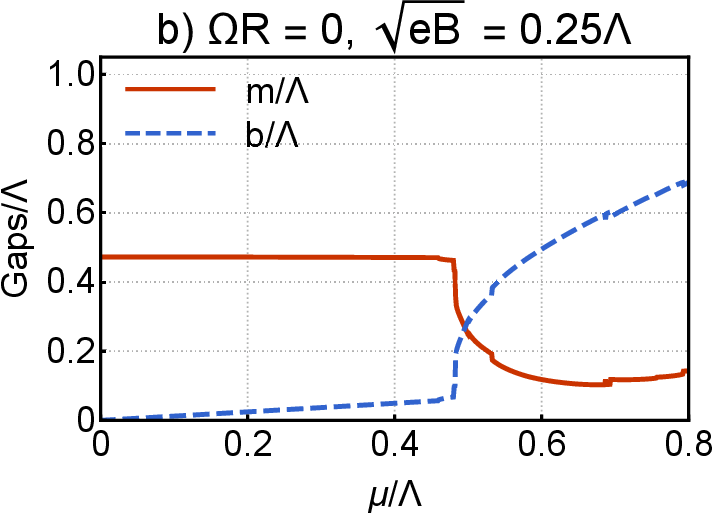}
	\includegraphics[width=5.5cm, height=4.5cm]{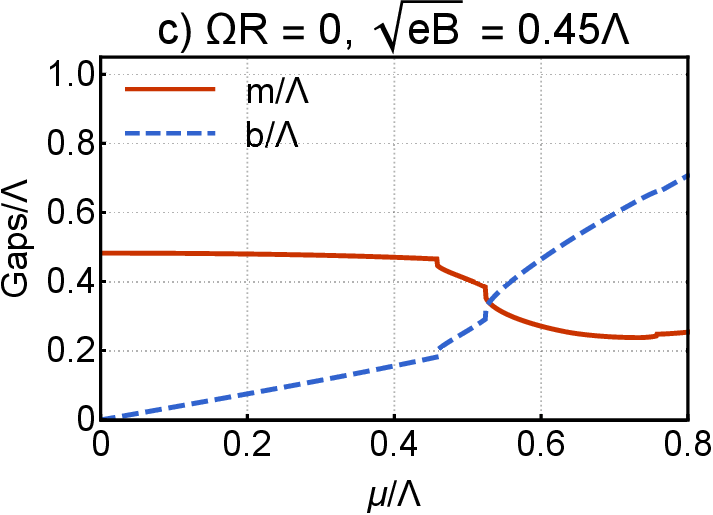}
	\includegraphics[width=5.5cm, height=4.5cm]{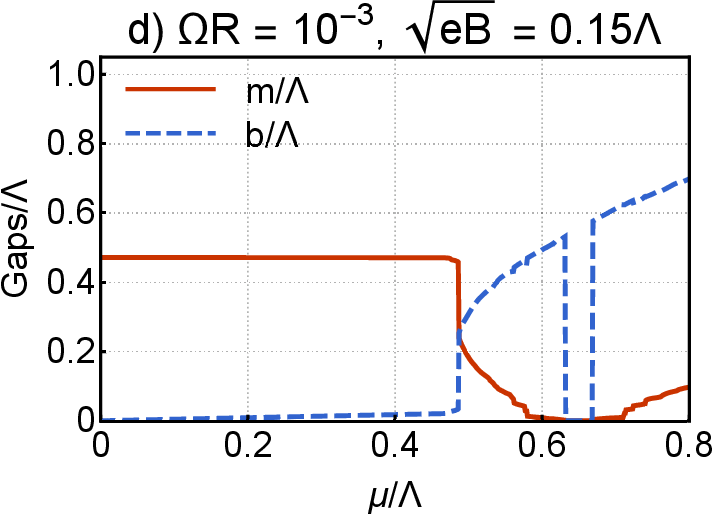}
	\includegraphics[width=5.5cm, height=4.5cm]{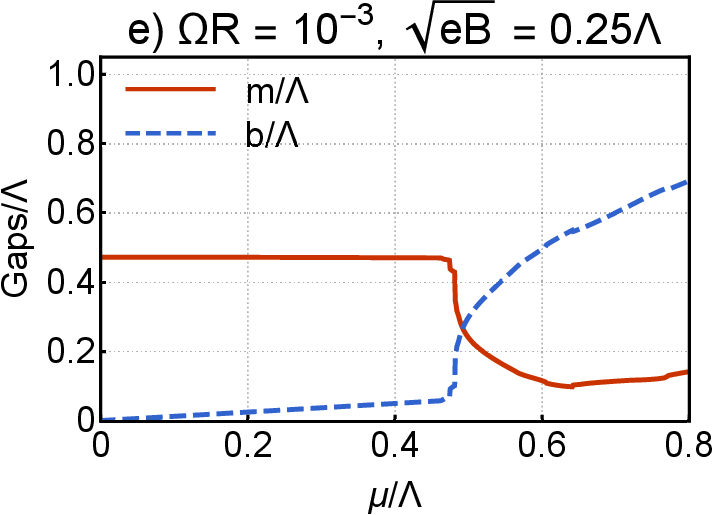}
	\includegraphics[width=5.5cm, height=4.5cm]{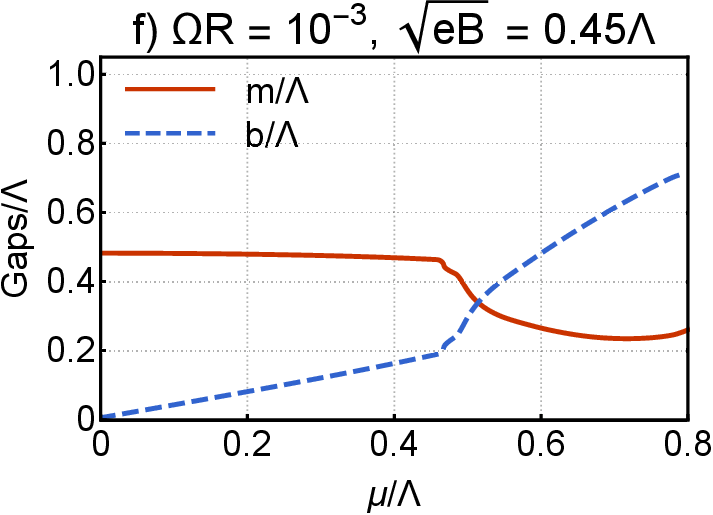}
	\includegraphics[width=5.5cm, height=4.5cm]{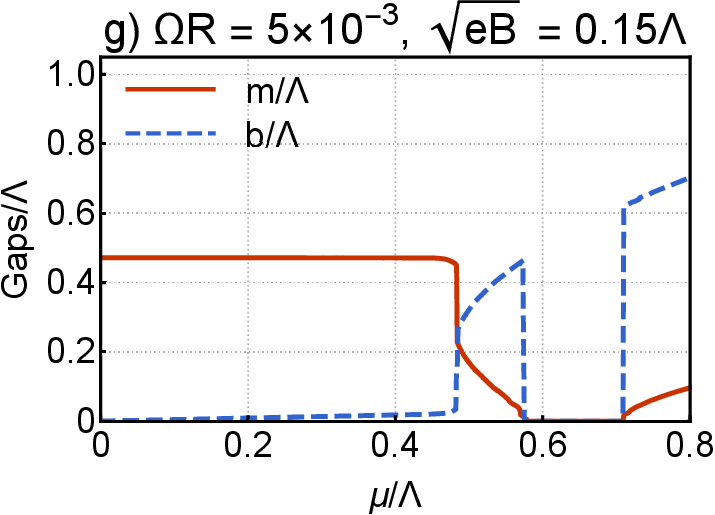}
	\includegraphics[width=5.5cm, height=4.5cm]{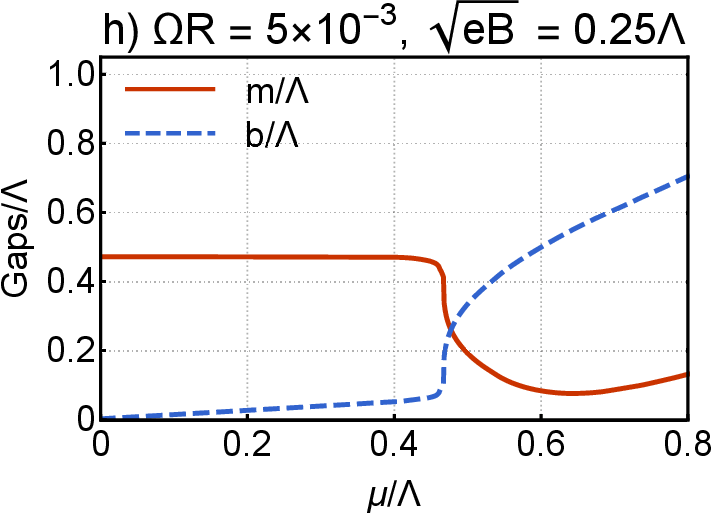}
	\includegraphics[width=5.5cm, height=4.5cm]{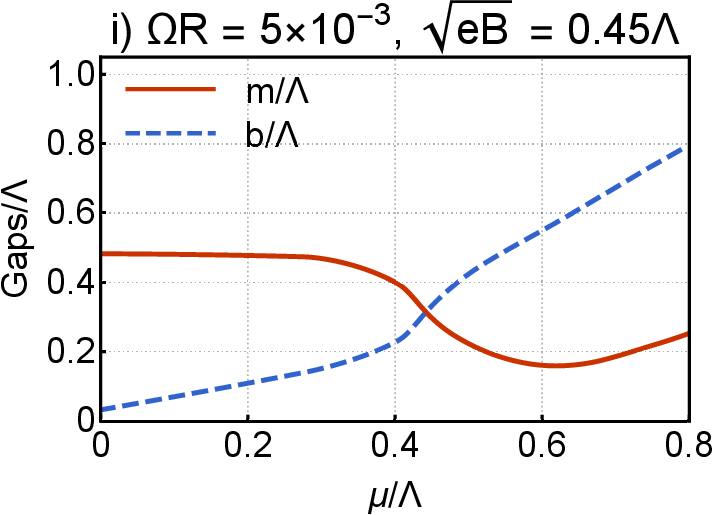}
	\includegraphics[width=5.5cm, height=4.5cm]{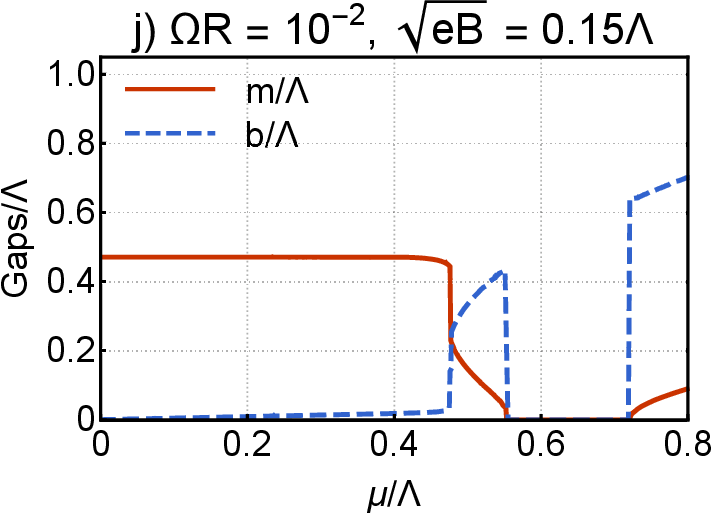}
	\includegraphics[width=5.5cm, height=4.5cm]{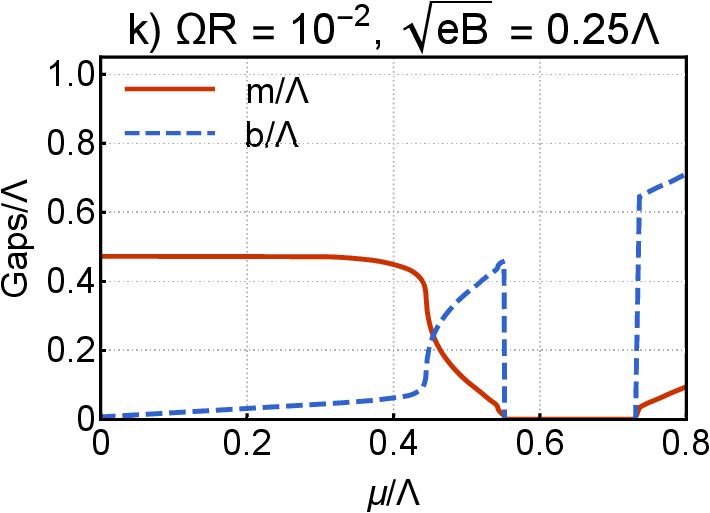}
	\includegraphics[width=5.5cm, height=4.5cm]{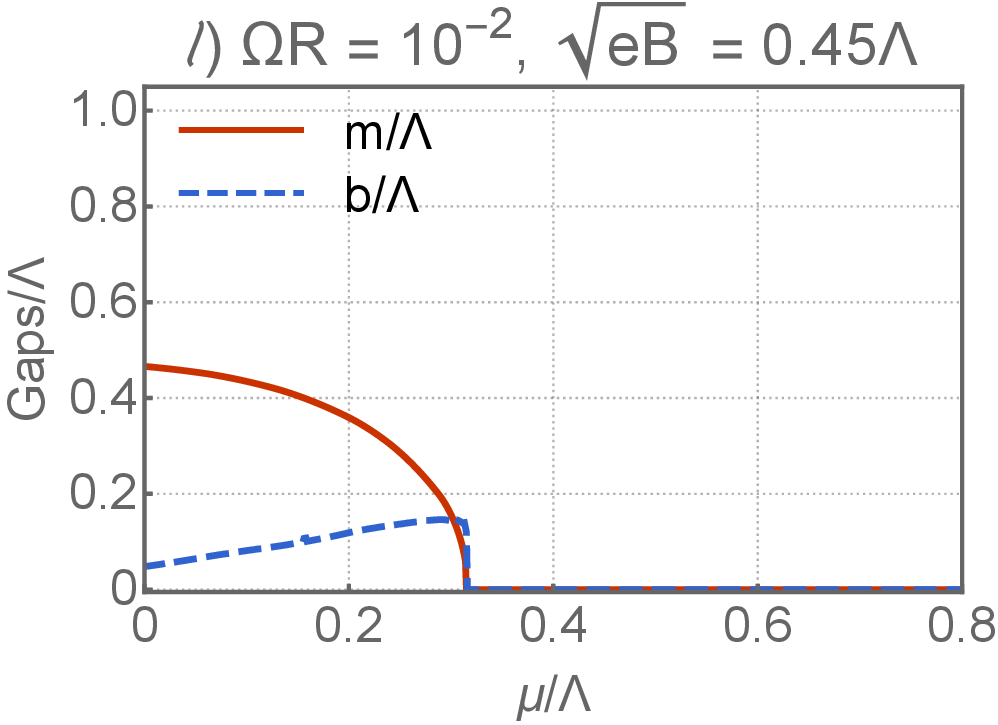}
\caption{color online. The $\mu$ dependence of the chiral condensate $m$ and spatial modulation $b$ are plotted for fixed values of $\Omega R$ and $\sqrt{eB}$. Solid orange and dashed blue curves correspond to $m$ and $b$, respectively.  In each row (column) $\Omega R$s are constant (vary) and $\sqrt{eB}$s vary (are constant). First row (panels a, b, and c): $\Omega R=0$ and $\sqrt{eB}=0.15\Lambda, 0.25\Lambda, 0.45\Lambda$. Second row (panels d, e, and f): $\Omega R=10^{-3}$ and $\sqrt{eB}=0.15\Lambda, 0.25\Lambda, 0.45\Lambda$. Third row (panels g, h, and i): $\Omega R=5\times 10^{-3}$ and $\sqrt{eB}=0.15\Lambda, 0.25\Lambda, 0.45\Lambda$. Fourth row (panels j, k, and l): $\Omega R=10^{-2}$ and $\sqrt{eB}=0.15\Lambda, 0.25\Lambda, 0.45\Lambda$.}\label{fig1}
\end{figure*}

\begin{figure*}[hbt]
	\includegraphics[width=5.5cm, height=4.5cm]{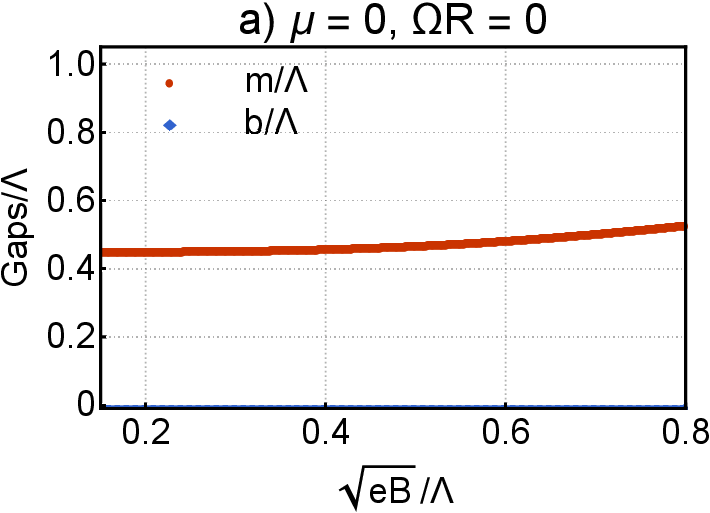}
	\includegraphics[width=5.5cm, height=4.5cm]{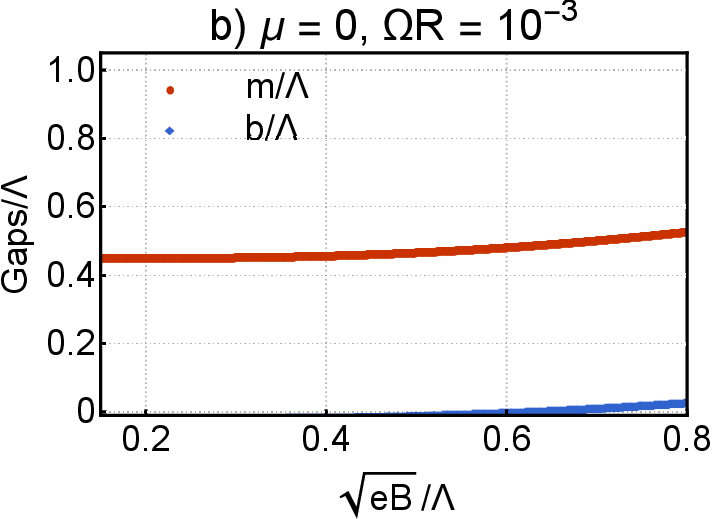}
	\includegraphics[width=5.5cm, height=4.5cm]{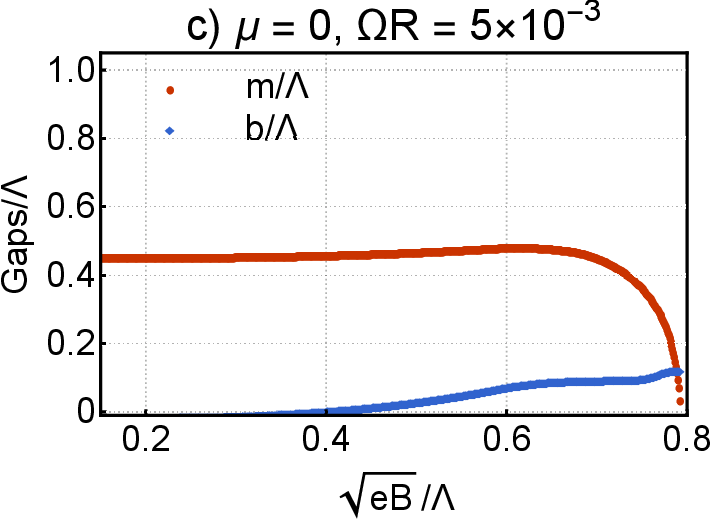}
	\includegraphics[width=5.5cm, height=4.5cm]{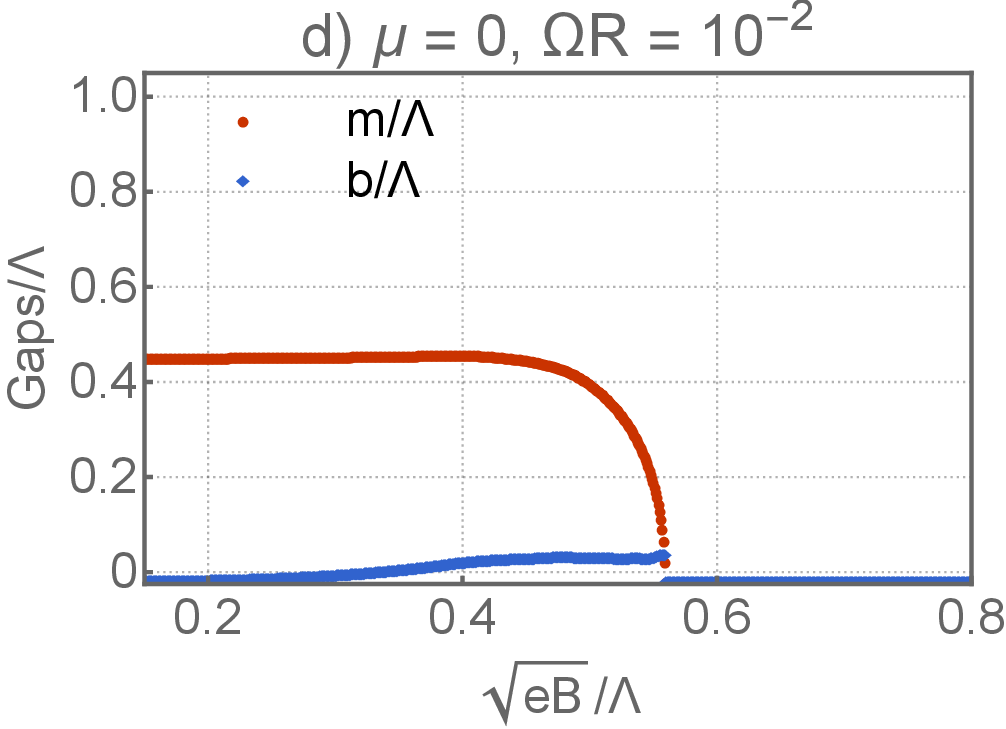}
	\includegraphics[width=5.5cm, height=4.5cm]{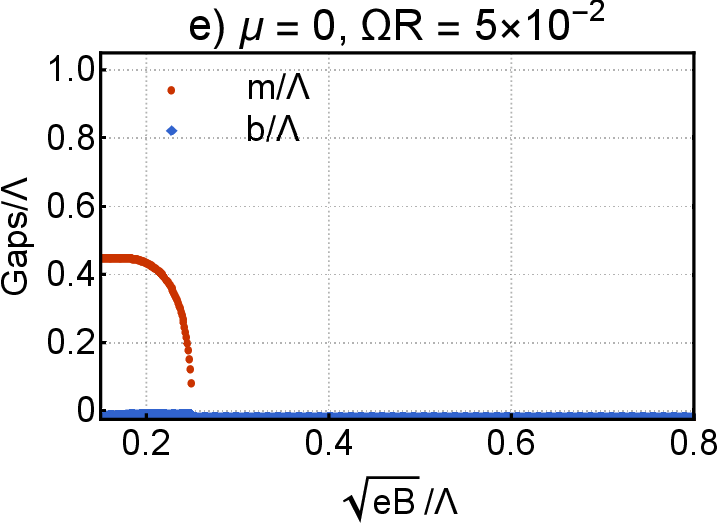}
	\caption{color online.The dependence of the chiral condensate $m$ and spatial modulation $b$ on $\sqrt{eB}$ is plotted for $\mu=0$ and $\Omega R=0,10^{-3},5\times 10^{-3},10^{-2}, 5\times 10^{-2}$. Orange dots and blue diamonds correspond to $m/\Lambda$ and $b/\Lambda$, respectively. It turns out that even for $\mu=0$, the interplay between $eB$ and $\Omega R$ leads to a finite $b$.}\label{fig2}
\end{figure*}

\begin{figure*}[hbt]
	\includegraphics[width=5.5cm, height=4.5cm]{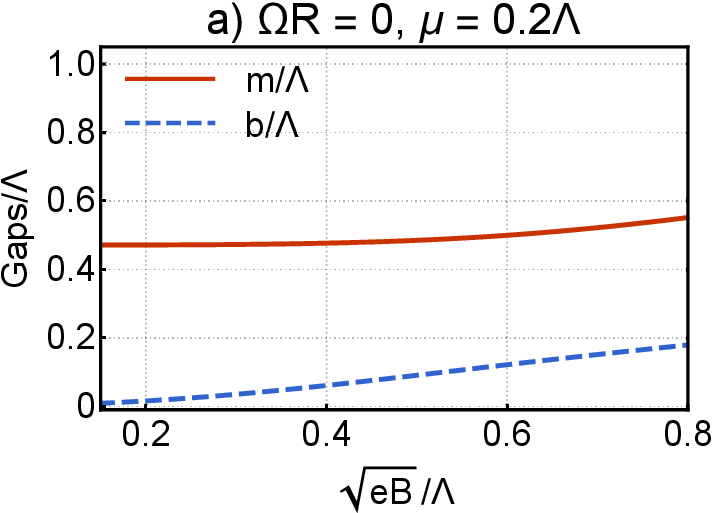}
	\includegraphics[width=5.5cm, height=4.5cm]{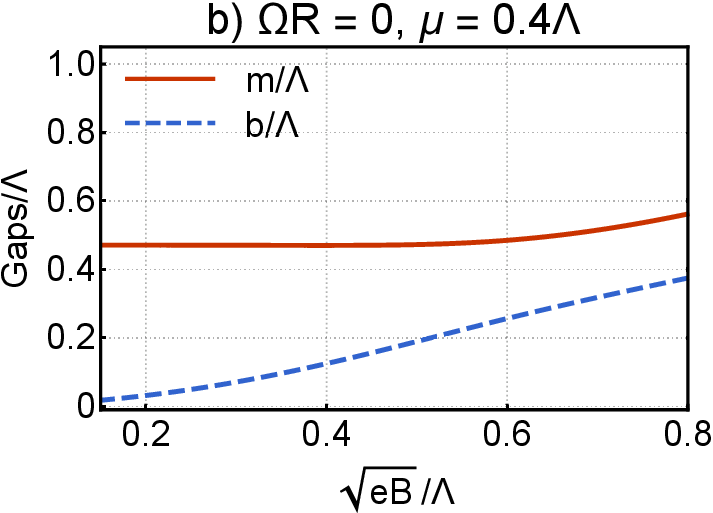}
	\includegraphics[width=5.5cm, height=4.5cm]{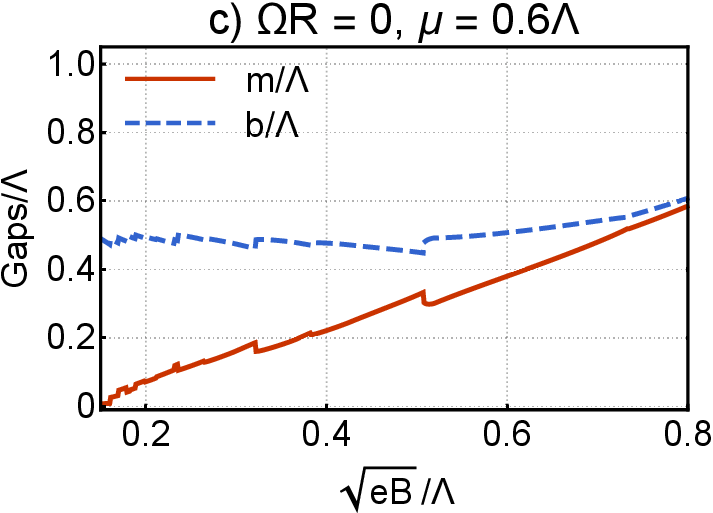}
	\includegraphics[width=5.5cm, height=4.5cm]{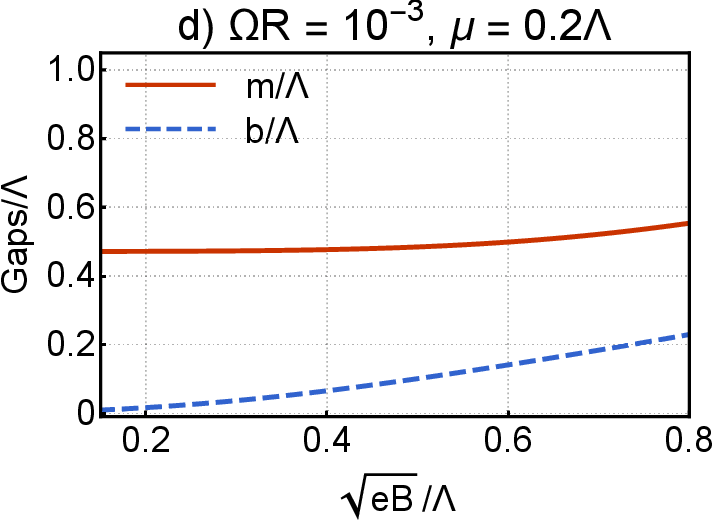}
	\includegraphics[width=5.5cm, height=4.5cm]{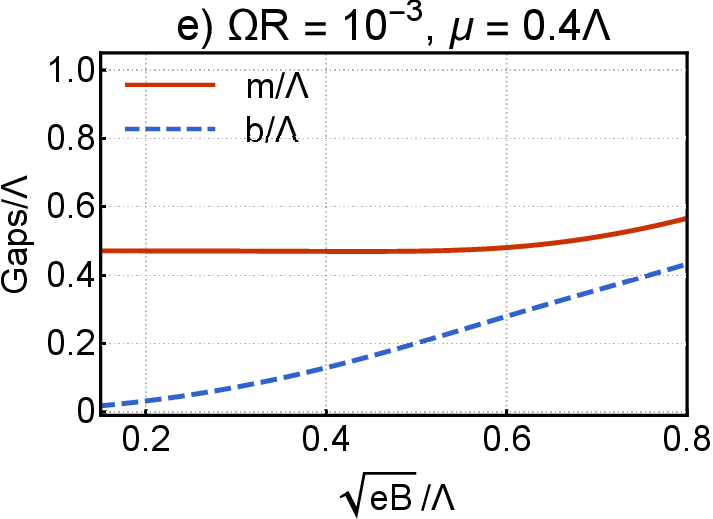}
	\includegraphics[width=5.5cm, height=4.5cm]{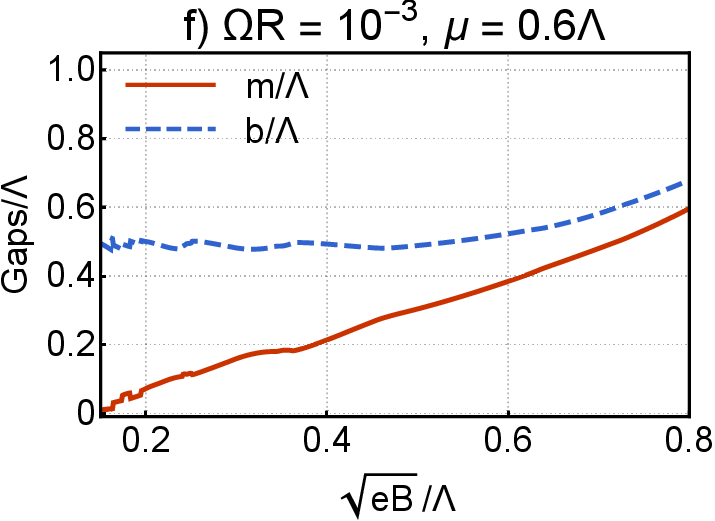}
	\includegraphics[width=5.5cm, height=4.5cm]{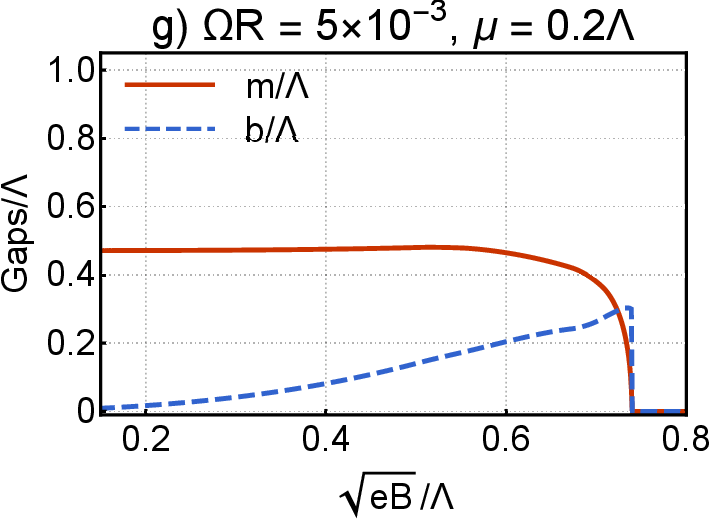}
	\includegraphics[width=5.5cm, height=4.5cm]{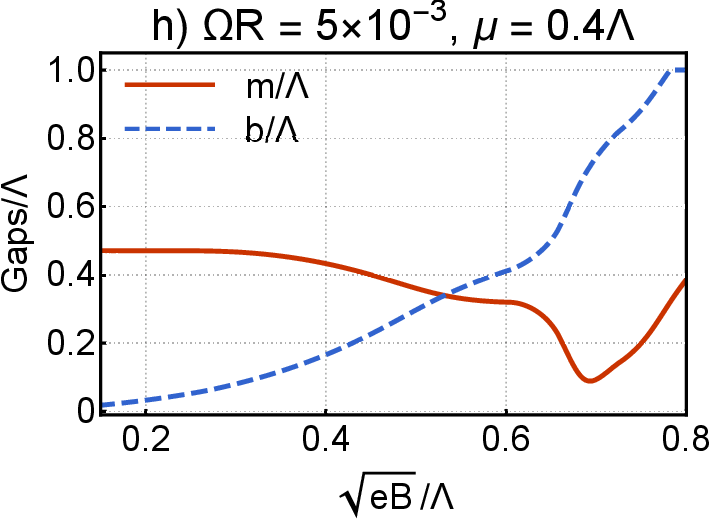}
	\includegraphics[width=5.5cm, height=4.5cm]{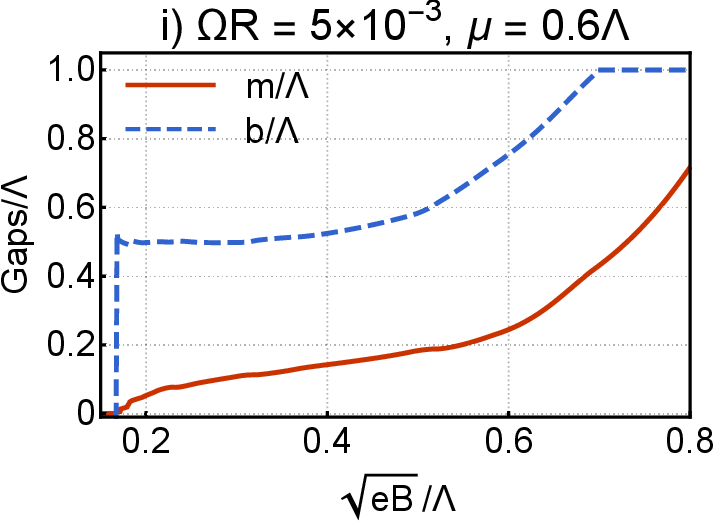}
	\includegraphics[width=5.5cm, height=4.5cm]{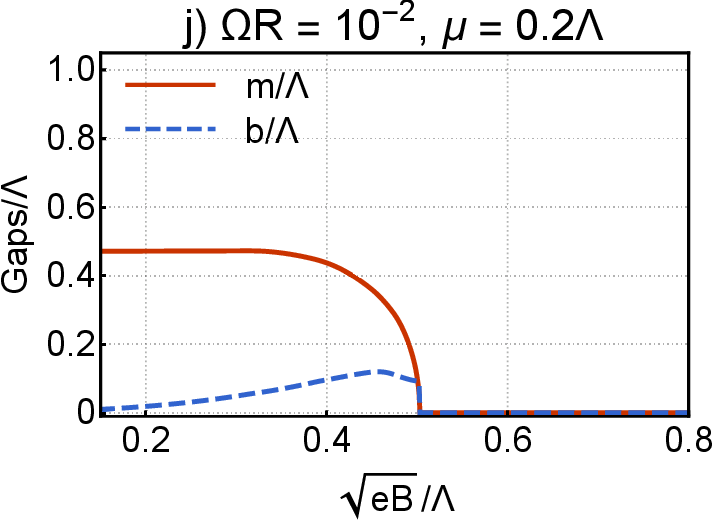}
	\includegraphics[width=5.5cm, height=4.5cm]{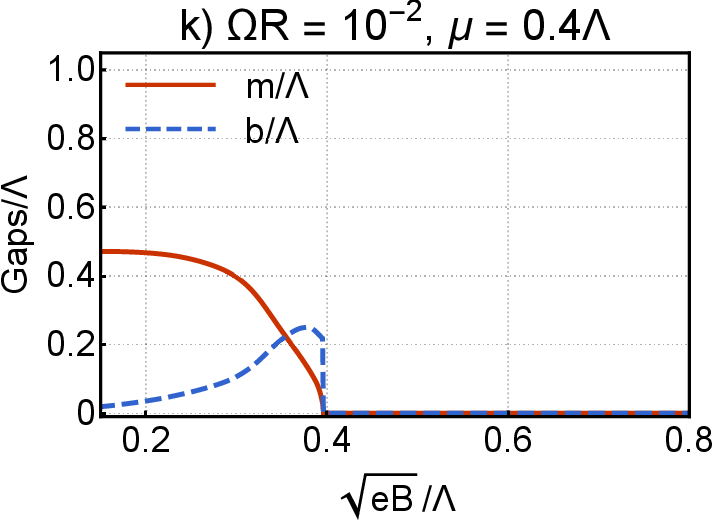}
	\includegraphics[width=5.5cm, height=4.5cm]{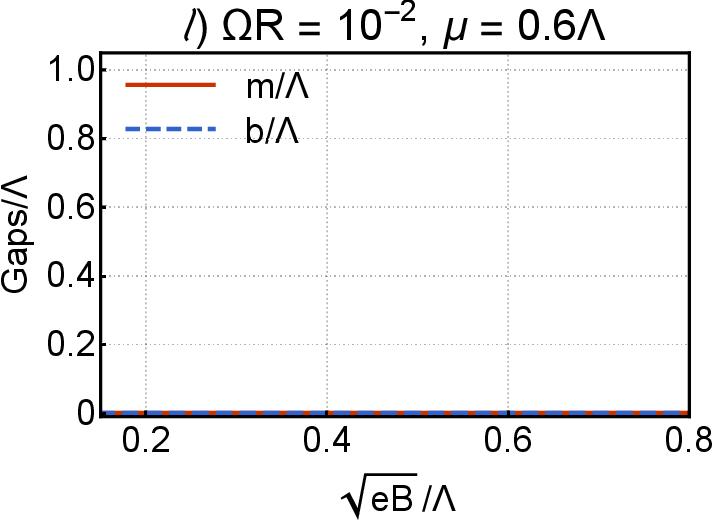}
	\caption{color online. The dependence of the chiral condensate $m$ and spatial modulation $b$ on $\sqrt{eB}$ is plotted for fixed values of $\Omega R$ and $\mu$. Solid orange and dashed blue curves correspond to $m$ and $b$, respectively. In each row (column) $\Omega R$s are constant (vary) and $\mu$s vary (are constant). First row (panels a, b, and c): $\Omega R=0$ and $\mu=0.2\Lambda, 0.4\Lambda, 0.6\Lambda$. Second row (panels d, e, and f): $\Omega R=10^{-3}$ and $\mu=0.2\Lambda, 0.4\Lambda, 0.6\Lambda$. Third row (panels g, h, and i): $\Omega R=5\times 10^{-3}$ and $\mu=0.2\Lambda, 0.4\Lambda, 0.6\Lambda$. Fourth row (panels j, k, and l): $\Omega R=10^{-2}$ and $\mu=0.2\Lambda, 0.4\Lambda, 0.6\Lambda$. For the plots to be comparable, the range in the horizontal axis is chosen to be in the interval $\sqrt{eB}\in [0.15\Lambda,0.8\Lambda]$.
}\label{fig3}
\end{figure*}

\begin{figure*}[hbt]
	\includegraphics[width=5.5cm, height=4.5cm]{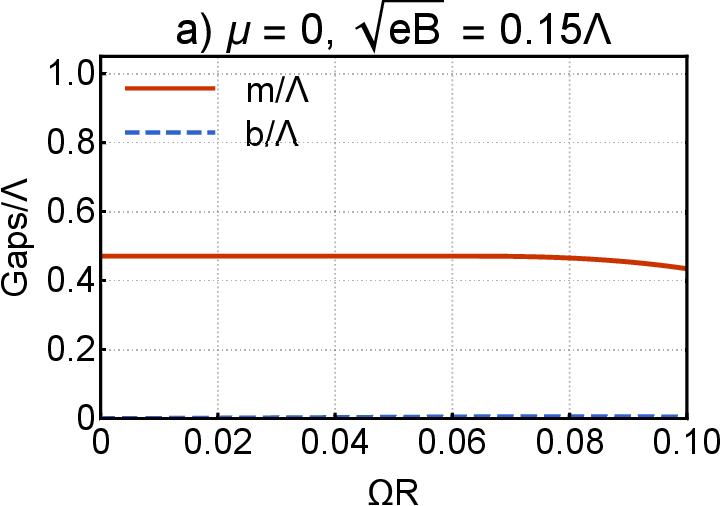}
	\includegraphics[width=5.5cm, height=4.5cm]{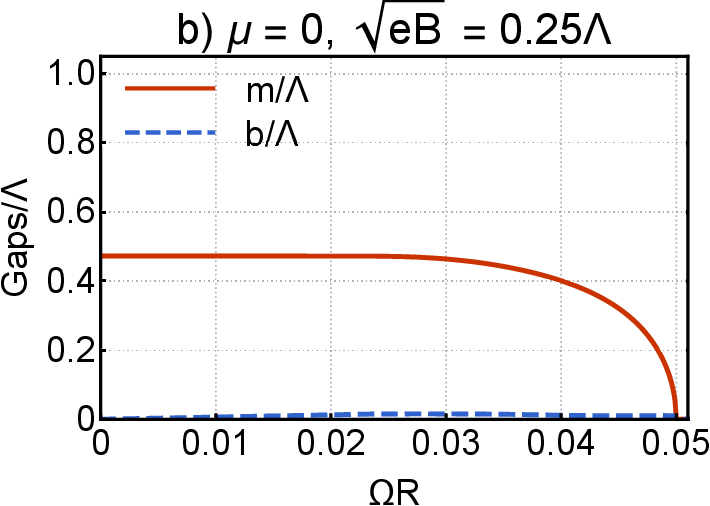}
	\includegraphics[width=5.5cm, height=4.5cm]{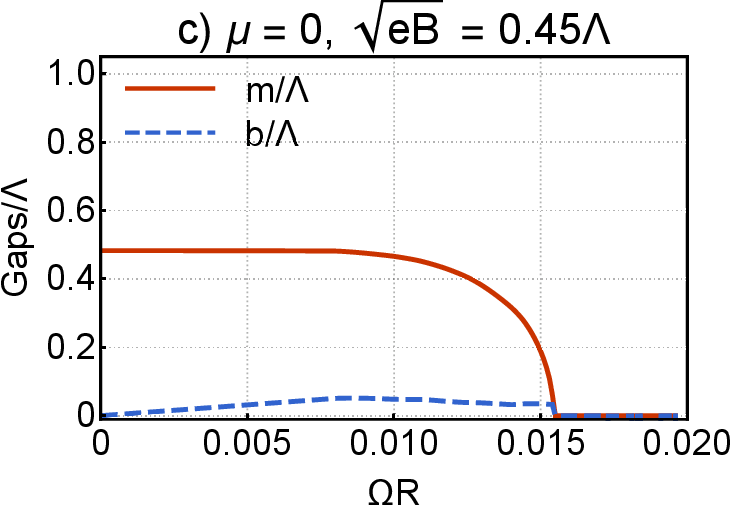}
	\includegraphics[width=5.5cm, height=4.5cm]{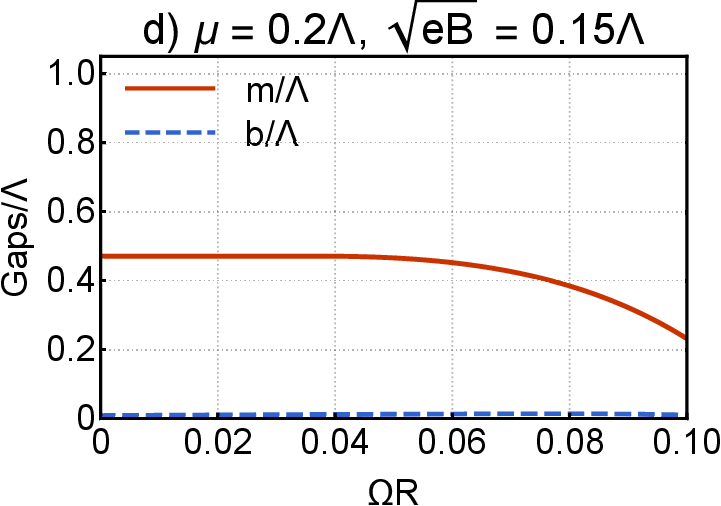}
	\includegraphics[width=5.5cm, height=4.5cm]{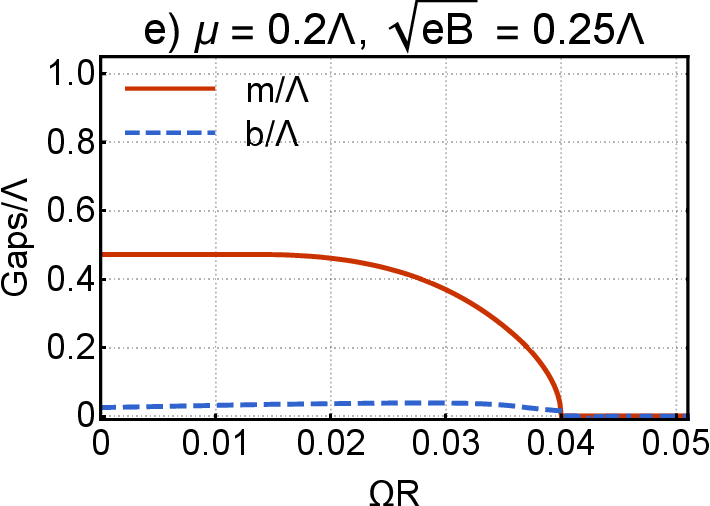}
	\includegraphics[width=5.5cm, height=4.5cm]{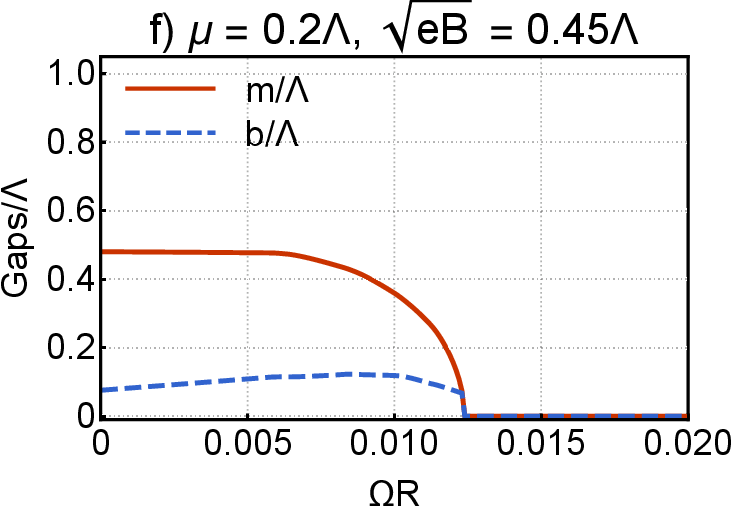}
	\includegraphics[width=5.5cm, height=4.5cm]{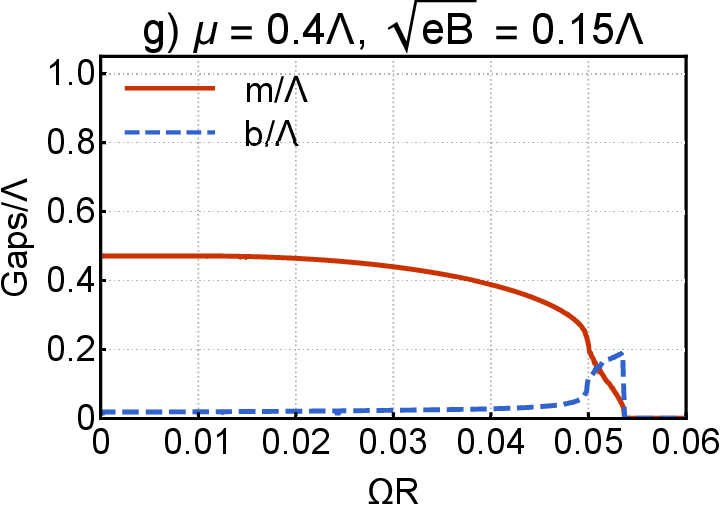}
	\includegraphics[width=5.5cm, height=4.5cm]{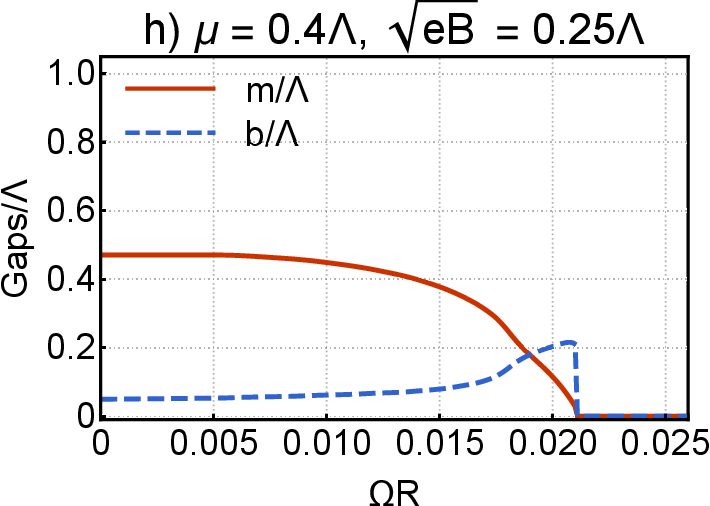}
	\includegraphics[width=5.5cm, height=4.5cm]{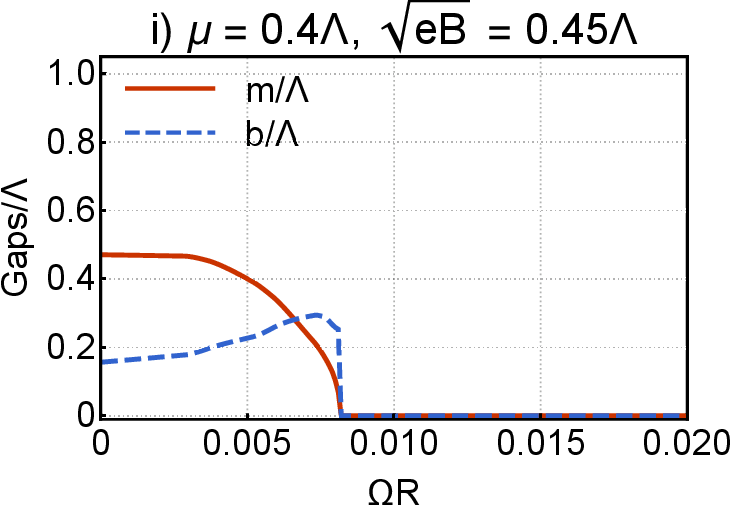}
	\includegraphics[width=5.5cm, height=4.5cm]{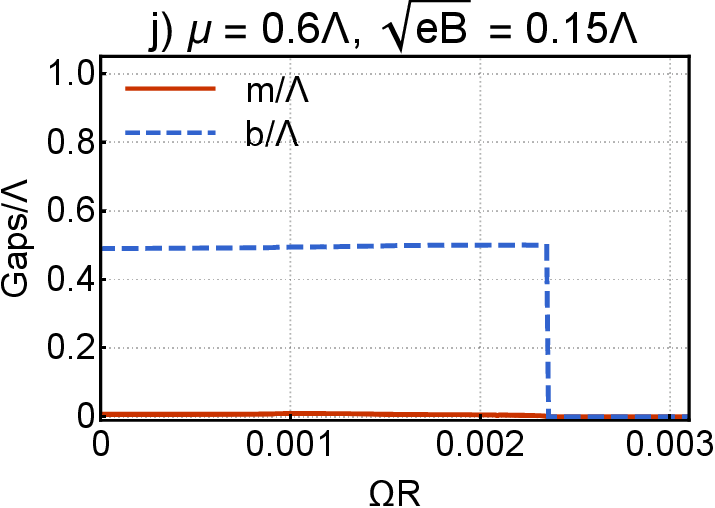}
	\includegraphics[width=5.5cm, height=4.5cm]{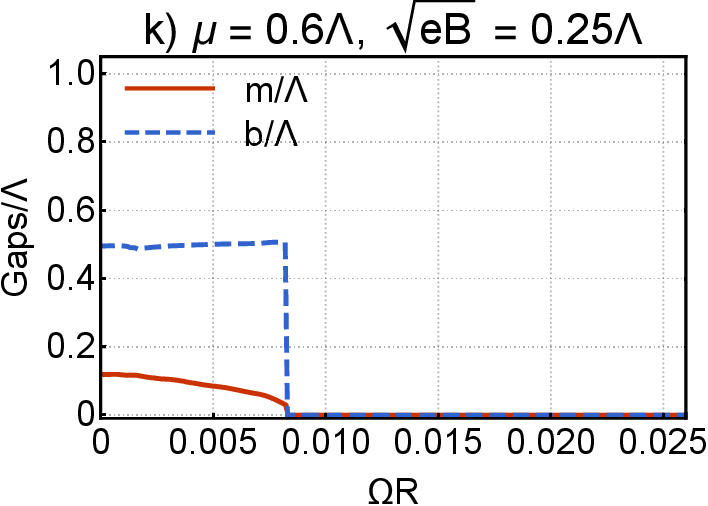}
	\includegraphics[width=5.5cm, height=4.5cm]{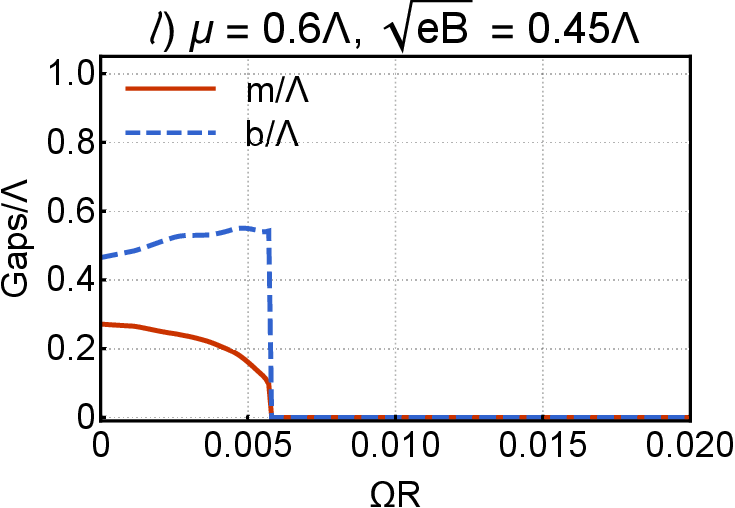}
	\caption{color online. The $\Omega R$ dependence of the chiral condensate $m$ and spatial modulation $b$ are plotted for fixed values of $\sqrt{eB}$ and $\mu$. Solid orange and dashed blue curves correspond to $m$ and $b$, respectively. First row (panels a, b, and c): $\mu=0$ and $\sqrt{eB}=0.15\Lambda, 0.25\Lambda, 0.45\Lambda$. Second row (panels d, e, and f): $\mu=0.2\Lambda$ and $\sqrt{eB}=0.15\Lambda, 0.25\Lambda, 0.45\Lambda$. Third row (panels g, h, and i): $\mu=0.4\Lambda$ and $\sqrt{eB}=0.15\Lambda, 0.25\Lambda, 0.45\Lambda$. Fourth row (panels j, k, and l): $\mu=0.6\Lambda$ and $\sqrt{eB}=0.15\Lambda, 0.25\Lambda, 0.45\Lambda$. }\label{fig4}
\end{figure*}
\begin{figure*}[hbt]
	\includegraphics[width=5.5cm, height=4.5cm]{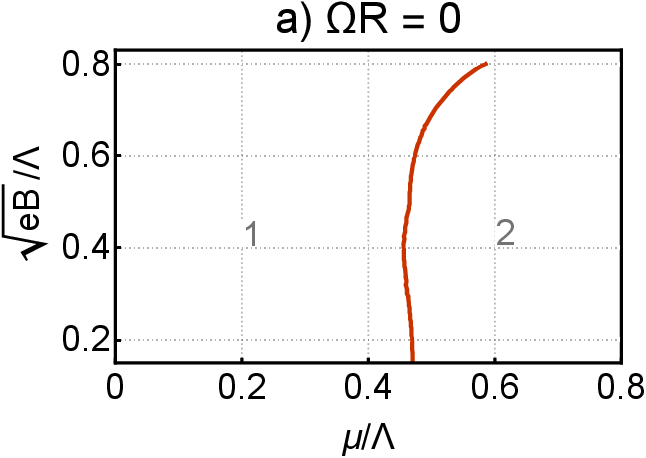}
	\includegraphics[width=5.5cm, height=4.5cm]{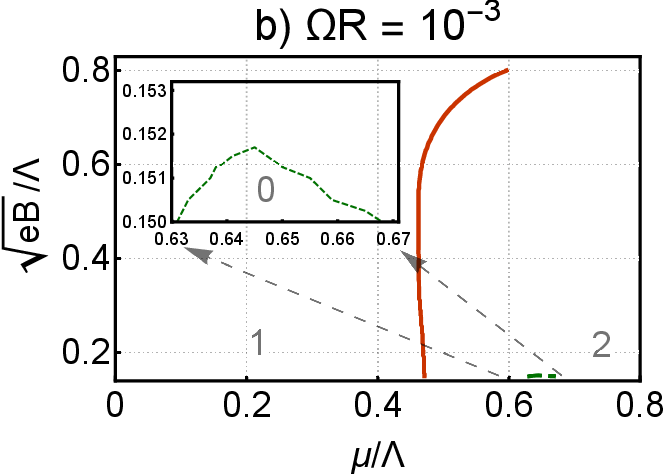}
	\includegraphics[width=5.5cm, height=4.5cm]{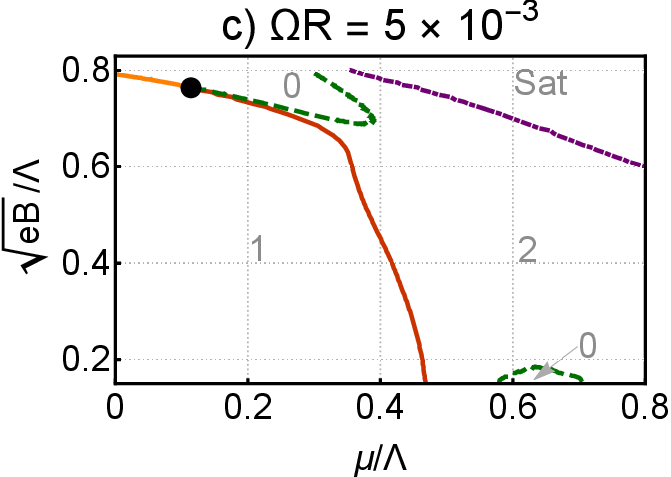}
	\includegraphics[width=5.5cm, height=4.5cm]{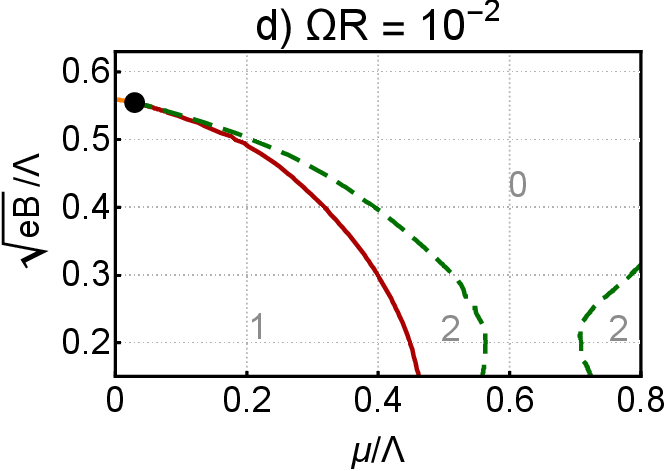}
	\includegraphics[width=5.5cm, height=4.5cm]{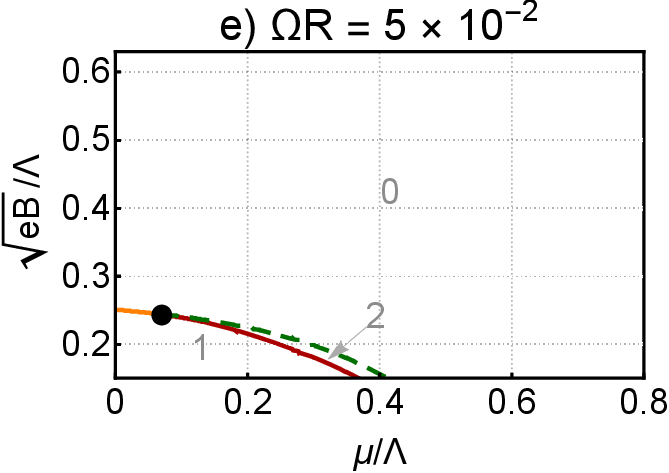}
\caption{color online. The phase portrait $\sqrt{eB}/\Lambda$ vs. $\mu/\Lambda$ for fixed $\Omega R=0, 10^{-3}, 5\times 10^{-3}, 10^{-2}, 5\times 10^{-2}$. The red solid lines separate two different regimes $\mu<m$ (denoted by "1") and $\mu>m$ (denoted by "2") in the \MD~phase, and the orange solid and green dashed lines separate the \MD~phase with $\mu<m$ and $\mu>m$ from the symmetry restored phase with $m=0$ (denoted by "0"). Black circles demonstrate the position of the critical points at  $(\mu,\sqrt{eB})\sim(0.115\Lambda,0.765\Lambda)$ (panel c), $(\mu,\sqrt{eB})\sim(0.03\Lambda,0.554\Lambda)$ (panel d) and $(\mu,\sqrt{eB})\sim(0.069\Lambda,0.243\Lambda)$ (panel e). The orange solid and green dashed lines correspond to first- and second-order phase transition lines, respectively. The regime denoted by "Sat", which is separated from the $\mu>m$ regime of the \MD~phase by a dotted-dashed magenta line, shall be excluded from the phase diagram (see the description in text). }\label{fig5}
\end{figure*}
\begin{figure*}[hbt]
	\includegraphics[width=5.5cm, height=4.5cm]{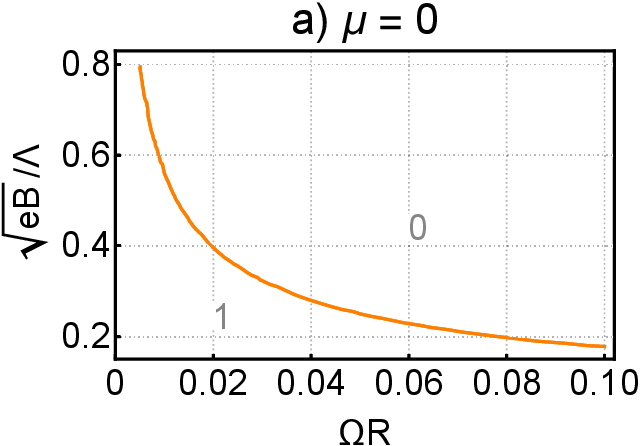}
	\includegraphics[width=5.5cm, height=4.5cm]{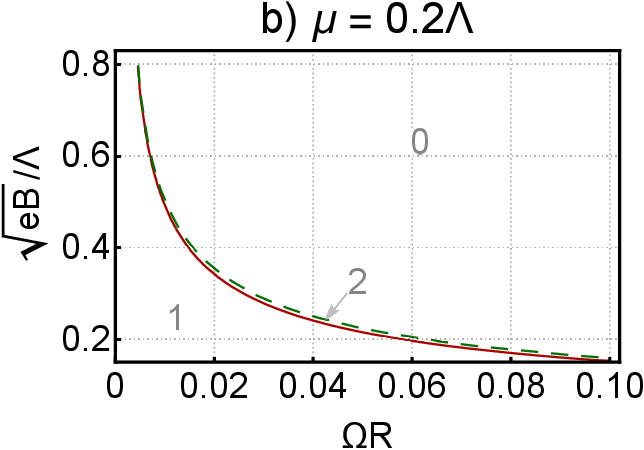}
	\includegraphics[width=5.5cm, height=4.5cm]{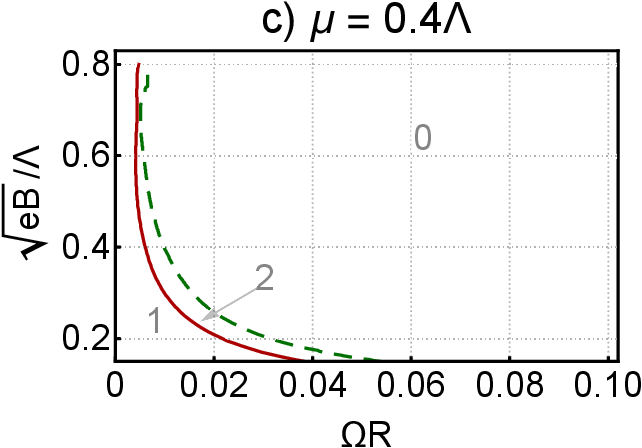}
	\includegraphics[width=5.5cm, height=4.5cm]{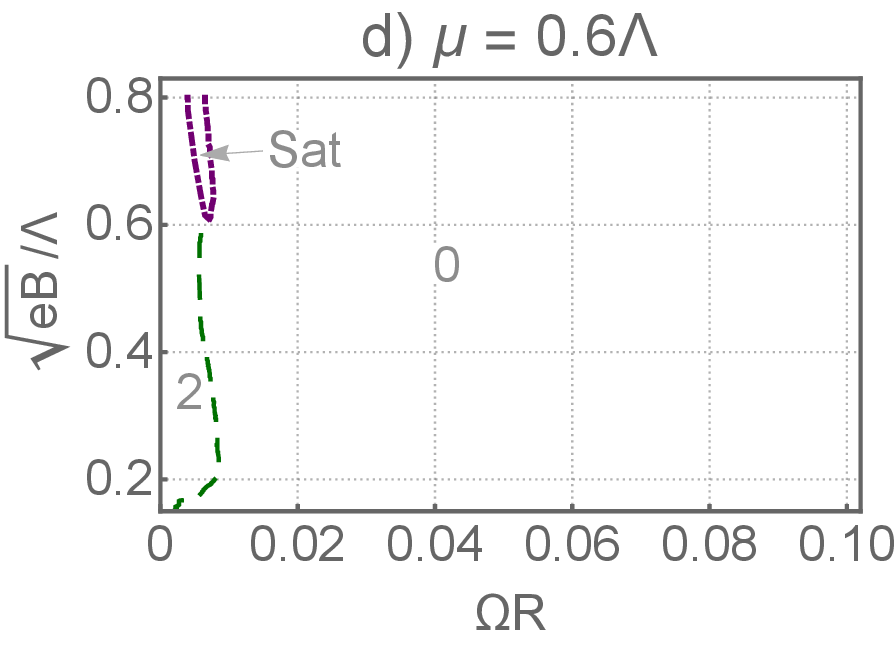}
\caption{color online. The phase portrait $\sqrt{eB}/\Lambda$ vs. $\Omega R$ for fixed $\mu/\Lambda=0,0.2,0.4,0.6$. The red solid lines separate two different regimes $\mu<m$ (denoted by "1") and $\mu>m$ (denoted by "2") in the \MD~phase, and the orange solid and green dashed lines separate the \MD~phase with $\mu<m$ and $\mu>m$ from the symmetry restored phase with $m=0$ (denoted by "0").  The orange solid line in panel (a) and the green dashed lines in panels (b)-(d) correspond to first- and second-order phase transitions. The regime denoted by "Sat", which is separated from the $\mu>m$ regime of the \MD~phase by a dotted-dashed magenta line, shall be excluded from the phase diagram (see the description in text). }\label{fig6}
\end{figure*}
\begin{figure*}[hbt]
	\includegraphics[width=5.5cm, height=4.5cm]{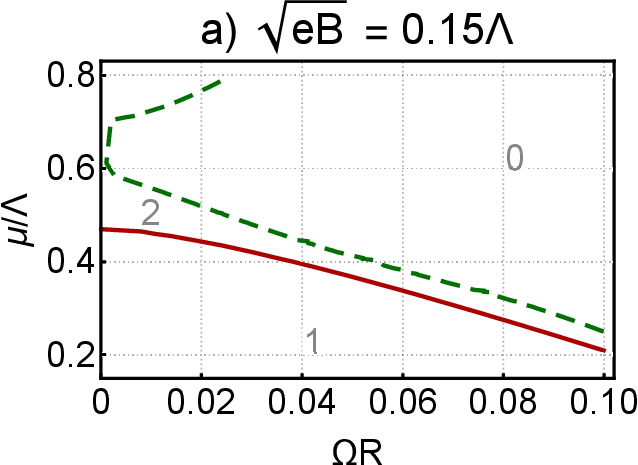}
	\includegraphics[width=5.5cm, height=4.5cm]{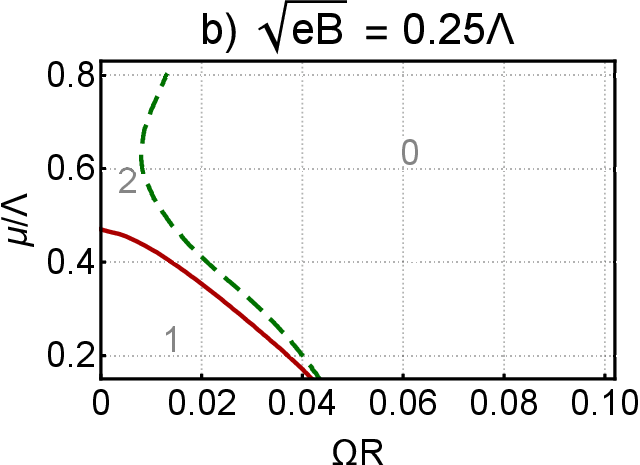}
	\includegraphics[width=5.5cm, height=4.5cm]{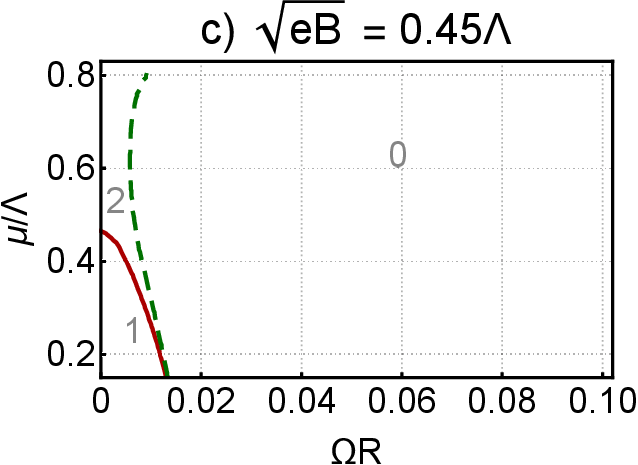}
\caption{color online. The phase portrait $\mu/\Lambda$ vs. $\Omega R$ for fixed $\sqrt{eB}/\Lambda=0.15, 0.25, 0.45$. The red solid lines separate two different regimes $\mu<m$ (denoted by "1") and $\mu>m$ (denoted by "2")  in the  \MD~phase, and the green dashed lines separate the  \MD~phase with $\mu>m$ from the symmetry restored phase with $m=0$ (denoted by "0"). Green dashed lines correspond to second-order phase transition lines. }\label{fig7}
\end{figure*}
\subsubsection{The $\mu$ dependence}
In Fig. \ref{fig1}, the $\mu$ dependence of the chiral condensate $m$ (solid orange) and spatial modulation $b$ (dashed blue) is plotted for fixed values of $\Omega R$ and $\sqrt{eB}$. In each row $\Omega R$ is constant and $\sqrt{eB}$ varies. In contrast, in each column $\sqrt{eB}$ is constant and  $\Omega R$ varies. The first to fourth rows correspond to $\Omega R=0,10^{-3},5\times 10^{-3}$ and $10^{-2}$, and the first to third columns correspond to $\sqrt{eB}=0.15\Lambda, 0.25\Lambda$ and $0.45\Lambda$. Let us first compare the plots in the first row. The result from Fig. \ref{fig1}(a) is comparable with the result presented in \cite{incera2022}. The four different regions described in \cite{incera2022} can also be identified in this plot. In the first region $\mu\lesssim 0.45\Lambda$, the chiral symmetry is broken, whereas the system is spatially symmetric ($b=0$). In the second region, $0.45\Lambda\lesssim \mu\lesssim 0.6\Lambda$, $m$ decreases, whereas $b$ increases.
The remnant mass is visible in a third region, $0.6\Lambda\lesssim \mu\lesssim 0.75\Lambda$. The spatial modulation $b$ is nonvanishing and rather large in this regime. A fourth region is visible in $\mu\gtrsim 0.75\Lambda$. In this regime $m$ and $b$ both increase with increasing $\mu$.
\footnote{Let us emphasize that Fig. \ref{fig1}(a) cannot be compared with Fig. 3(b) of \cite{frolov2010}, because as it is argued in \cite{incera2022}, a very small remnant mass $m\sim 5$ MeV in the intermediate regime of densities $0.6\Lambda\lesssim\mu\lesssim 0.75\Lambda$ is not considered in \cite{frolov2010}, and, consequently, $b$ is reported to be zero in this regime.}
These four regimes are visible in all the plots in the first column [Figs. \ref{fig1}(a),  \ref{fig1}(d),  \ref{fig1}(g), and  \ref{fig1}(j)], where $\sqrt{eB}=0.15\Lambda$ is constant, and $\Omega R$ increases. Here, $\Omega R$ merely affects the production of $m$. It decreases (vanishes) by increasing $\Omega R$. As it turns out, the remnant mass appearing in the third region, $0.6\Lambda\lesssim \mu\lesssim 0.75\Lambda$, disappears. Consequently, $b$ also vanishes in this regime. This specific feature of $\Omega R$ in destroying the dynamical mass is expected from \cite{fukushima2015, sadooghi2021}. As concerns the fourth regime, however, it turns out that the system reenters from a chirally symmetric homogeneous regime into a chiral symmetry broken phase for $\mu\gtrsim 0.75\Lambda$, where the spatial symmetry is also broken by a nonvanishing and rather large $b$.
\par
By increasing $\sqrt{eB}$ and for small values of $\Omega R$, the remnant mass in the third region increases. This is expected from the magnetic catalysis, and is visible by comparing the plots in the second column [Figs. \ref{fig1}(b),  \ref{fig1}(e),  \ref{fig1}(h)], corresponding to $\sqrt{eB}=0.25\Lambda$ with the ones in the third column [Figs. \ref{fig1}(c),  \ref{fig1}(f),  \ref{fig1}(i)], corresponding to $\sqrt{eB}=0.45\Lambda$. Here, as it turns out, $b$ increases also in the first region $\mu\lesssim 0.4\Lambda$. However, once $\Omega R$ is large enough, $m$ vanishes in the third region because of the IMRC \cite{sadooghi2021}. Larger $\Omega R$ destroys the dynamical mass even in the fourth regime $\mu\gtrsim 0.75\Lambda$ [see Fig. \ref{fig1}($\ell$)] with $\Omega R=10^{-2}$ and $\sqrt{eB}=0.45\Lambda$. Hence, large enough $\Omega R$ does not allow any reentrance in the symmetry broken phase in the fourth regime $\mu\gtrsim 0.75\Lambda$. In \ref{sec4B}, we study the effect of $eB$ and $\Omega R$ on the critical values of $\mu$.
\subsubsection{The $eB$ dependence}
In Fig. \ref{fig2}, the $eB$ dependence of the chiral condensate $m$ (orange dots) and spatial modulation $b$ (blue diamonds) is plotted for $\mu=0$ and $\Omega R=0,10^{-3}, 5\times 10^{-3},10^{-2},5\times 10^{-2}$. As expected for $\mu=0$ and $\Omega R=0$, the dynamical mass increases slightly with increasing $eB$, and $b=0$ [Fig. \ref{fig2}(a)]. For a nonvanishing $\Omega R=0.001$, however, $b$ increases slightly with increasing $eB$ [Fig. \ref{fig2}(b)], and for $\Omega R=0.005$, because of the onset of the IMRC effect, $m$ decreases in the regime $\sqrt{eB}\gtrsim 0.7\Lambda$. The spatial modulation $b$, however, increases with increasing $eB$ [see Fig. \ref{fig2}(c)].
We conclude therefore that even for $\mu=0$, a certain interplay between $eB$ and $\Omega R$ leads to a finite $b$. This is indeed expected because of the similarity between $\mu$ and $\Omega R$ discussed in \cite{fukushima2015}.
\par
For larger values of $\Omega R$, the rotation completely destroys $m$, and for vanishing $m$, the spatial modulation $b$ also vanishes [Figs. \ref{fig2}(d) and \ref{fig2}(e)]. For $\Omega R\gtrsim 0.005$, the transition from the \MD~phase with $m,b\neq 0$ to the symmetry restored phase with $m=0$ is of first order [see Fig. \ref{fig6}(a)]. By comparing the plots from Figs. \ref{fig2}(c)-\ref{fig2}(e), it turns out that the critical value of $eB$, for which $m$ and $b$ vanish, decreases with increasing $\Omega R$. This result coincides with the results presented in the phase diagram $eB$-$\Omega R$ in Fig. \ref{fig6}.
\par
To study the effect of $\mu$ on the above scenario, the $eB$ dependence of the chiral condensate $m$ (solid orange) and spatial modulation $b$ (dashed blue) is plotted for fixed values of $\Omega R$ and nonvanishing $\mu$ in Fig. \ref{fig3}. In each row $\Omega R$ is constant and $\mu$ varies.  In contrast, in each column $\mu$ is constant and $\Omega R$ varies. The first to fourth rows correspond to $\Omega R=0,10^{-3},5\times 10^{-3}$ and $10^{-2}$, and the first to third columns correspond to $\mu=0.2\Lambda, 0.4\Lambda,$ and $0.6\Lambda$.
It is possible to compare the results presented in Figs. \ref{fig3}(a)-\ref{fig3}(c) with those from Fig. 2 of \cite{frolov2010}, where $\Omega R=0$. To do this, one shall, however, be very cautious because, according to \cite{incera2022} and also our results from Fig. \ref{fig1}, the small remnant mass at intermediate $\mu$ is not considered in \cite{frolov2010}. Hence, by comparing, e.g. Fig. \ref{fig3}(c) for $\Omega R=0$ and $\mu=0.6\Lambda$ with Fig. 2(d) in \cite{frolov2010}, we observe that a remnant mass $m\sim 5-70$ MeV appears in the regime $\sqrt{eB}\leq 0.2\Lambda$ of Fig. \ref{fig3}(c), while this mass is not considered in Fig. 2(d) of \cite{frolov2010} in the same regime of $\sqrt{eB}\leq 0.2\Lambda$. Let us emphasize that the result from Fig. \ref{fig3}(c) is consistent with Fig. \ref{fig1}(a) for $\Omega R=0$, which is by itself in complete agreement with the result presented in \cite{incera2022}.
\par
The plots presented in the first column of Fig. \ref{fig3} [Figs. \ref{fig3}(a), \ref{fig3}(d), \ref{fig3}(g), and \ref{fig3}(j)], corresponding to $\mu=0.2\Lambda$ (small $\mu$) and $\Omega R=0, 10^{-3}, 5\times 10^{-3}, 10^{-2}$ are quite similar to the plots from Fig. \ref{fig2}
corresponding to $\mu=0$ and the same $\Omega R$s. The onset of IMRC effect on destroying $m$ is visible for larger values of $\Omega R=5\times 10^{-3}, 10^{-2}$. Comparing \ref{fig2}(c) with \ref{fig3}(g), and \ref{fig2}(d) with \ref{fig3}(j), it turns out that larger $\mu$ reinforces the formation of $b$ [the slope of $b$ in \ref{fig3}(g) and  \ref{fig3}(j) are larger than the slope of $b$ in \ref{fig2}(c) and \ref{fig2}(d)]. The critical value of $eB$, for which $m$ and $b$ vanish, is, however, smaller for nonvanishing $\mu$.
\par
For moderate values of $\mu=0.4\Lambda$ and $\Omega R=0,10^{-3}$, the scenario is similar to the case of $\mu=0.2\Lambda$ with the same $\Omega R$'s [compare Figs. \ref{fig3}(b) and \ref{fig3}(e) with \ref{fig3}(a) and \ref{fig3}(d)]. For $\Omega R=5\times 10^{-3}$ and $\mu=0.4\Lambda$, however, $m$ decreases with increasing $eB$, whereas $b$ increases rapidly with increasing $eB$ [see Fig.\ref{fig3}(h)]. The fact that $m$ decreases for strong magnetic fields, moderate $\mu$, and relatively large $\Omega R$ is related to the IMRC effect \cite{sadooghi2021}. This effect becomes more apparent for larger values of $\Omega R=0.01$, which is large enough to destroy $m$, and consequently $b$, in the regime $\sqrt{eB}\gtrsim 0.4 \Lambda$ [see Fig. \ref{fig3}(k)].
\par
This scenario completely changes for larger values of $\mu$ (see the plots in the third column of Fig. \ref{fig3}). Here, for $\mu=0.6\Lambda$, $b$ is larger than $m$, as long as $\Omega R \lesssim 0.005$ [see \ref{fig3}(c), \ref{fig3}(f), and \ref{fig3}(i)]. For large value of $\Omega R=0.01$, both $m$ and $b$ vanish in the whole interval $\sqrt{eB}\in [0.15\Lambda,0.8\Lambda]$ [see Fig. \ref{fig3}($\ell$)]. This result coincides with the results presented in Figs. \ref{fig1}(j)-\ref{fig1}($\ell$). Comparing the plots from Figs. \ref{fig3}(j)-\ref{fig3}($\ell$), it turns out that for large $\Omega=0.01$, the critical value of $eB$ decreases with increasing $\mu$. This is expected from the IMRC effect \cite{sadooghi2021}, and is completely visualized in these plots.
\subsubsection{The $\Omega R$ dependence}
In Fig. \ref{fig4}, the $\Omega R$ dependence of the chiral condensate $m$ (solid orange) and spatial modulation $b$ (dashed blue) is plotted for fixed values of $\mu$ and $\sqrt{eB}$. In each row $\mu$ is constant and $\sqrt{eB}$ varies. In contrast, in each column $\sqrt{eB}$ is constant and  $\mu$ varies. The first to fourth rows correspond to $\mu =0,0.2\Lambda,0.4\Lambda,0.6\Lambda$, and the first to third columns correspond to $\sqrt{eB}=0.15\Lambda, 0.25\Lambda$ and $0.45\Lambda$. We notice that in contrast to previous figures, the ranges of the horizontal axes are not the same in all the plots from Fig. \ref{fig4}. Let us first compare the plots in each row: By comparing the plots from the first to third row [Figs. \ref{fig4}(a)-\ref{fig4}(c), \ref{fig4}(d)-\ref{fig4}(f), \ref{fig4}(g)-\ref{fig4}(i)] together, it turns out that the interplay between the IMRC effect arising from a simultaneous increase of $eB$ and $\Omega R$ leads to decreasing the critical value of $\Omega R$ with increasing $eB$. As concerns Figs. \ref{fig4}(j)-\ref{fig4}($\ell$) in the fourth row, however, large value of $\mu=0.6\Lambda$ leads to $b>m$, in contrast to all the other plots. In Fig. \ref{fig4}(j), a remnant mass ($m\neq 0$) in the regime $\Omega R\lesssim 0.002$, similar to that which appeared in the third region of Fig. \ref{fig1}(a), leads to a relatively large $b$ in this regime. By increasing $\sqrt{eB}$ this remnant mass becomes larger, as expected from magnetic catalysis, and,
at the same time, the critical $\Omega R$ decreases [compare Fig. \ref{fig4}(k) with \ref{fig4}($\ell$)]. The latter can be regarded as a signature of IMRC effect.
\subsection{Phase portraits $\boldsymbol{eB}$-$\boldsymbol{\mu}$, $\boldsymbol{eB}$-$\boldsymbol{\Omega R}$, and $\boldsymbol{\mu}$-$\boldsymbol{\Omega R}$}\label{sec4B}
In this section, we present the phase portraits $\sqrt{eB}$-$\mu$ for fixed $\Omega R$, $\sqrt{eB}$-$\Omega R$ for fixed $\mu$, and $\mu$-$\Omega R$ for fixed $\sqrt{eB}$, and demonstrate the impact of the IMRC effect on the phase diagrams of this model, as well as similarities/differences between $\mu$ and $\Omega R$ in creating the \MD~phase.
\par
In Fig. \ref{fig5}, the phase portrait $\sqrt{eB}/\Lambda$-$\mu/\Lambda$ for fixed $\Omega R=0, 10^{-3}, 5\times 10^{-3}, 10^{-2}, 5\times 10^{-2}$ is plotted. The red solid lines separate two different regimes $\mu<m$ (denoted by "1") and $\mu>m$ (denoted by "2") in the \MD~phase, orange solid and green dashed lines separate the \MD~phase with $\mu<m$ and $\mu>m$ from the symmetry restored phase with $m=0$ (this normal phase is denoted by "0"). They correspond to first- and second-order phase transition lines, respectively. Figure \ref{fig5}(a) is comparable with the results presented in \cite{frolov2010}, where only two regimes $\mu<m$ and $\mu>m$ appear in the whole parameter space. For $\Omega R=10^{-3}$, the parameter space is slightly different: A very small region of the normal phase appears in the interval $\mu {\in} [\sim 0.63\Lambda ,\sim 0.67\Lambda]$  for small $\sqrt{eB}\lesssim 0.152\Lambda$ [see the sub-figure inserted into Fig. \ref{fig5}(b)]. The phase transition from the $\mu>m$ regime of the \MD~phase to the normal phase is of second-order. This symmetry restored phase corresponds to the third region, which is demonstrated in Fig. \ref{fig1}(d). The $\sqrt{eB}/\Lambda$-$\mu/\Lambda$ phase diagram for $\Omega R=0.005$ in Fig.\ref{fig5}(c) becomes more complex. Apart from two regimes $\mu<m$ and $\mu>m$ that are indicated by the red solid line, and the orange solid and green dashed lines that separate the
\MD~phase with $\mu<m$ and $\mu>m$ from the normal phase, there is another region in the upper right corner of the phase portrait which is separated by a magenta dotted-dashed line from the $\mu>m$ regime of the \MD~phase. In this region, $b$ is larger than $\Lambda=1$ GeV, the cutoff of the model. This regime shall thus be completely excluded from the phase portrait. The black circle at $(\mu,\sqrt{eB})\sim (0.115\Lambda,0.795\Lambda)$ demonstrates the position of a critical point at the end of a second-order transition line.\footnote{The point is not a tricritical point because, according to the above description, the two regimes $\mu<m$ and $\mu>m$ are two distinct "regimes" in one and the same chirally broken \MD~phase. }
\par
In Fig. \ref{fig5}(d) for $\Omega R=0.01$, the phase portrait includes two regimes $\mu<m$ and $\mu>m$ in the \MD~phase. Because of the small $\mu>m$ regime in the lower right corner of the phase portrait, for $\sqrt{eB}\lesssim 0.3 \Lambda$, a reentrance from the normal phase into the \MD~ phase occurs for $\mu\gtrsim 0.75\Lambda$. For larger values of $\Omega R$s, the normal phase becomes larger and the \MD~phase becomes suppressed. As it is demonstrated in Fig. \ref{fig5}(e), for $\Omega R=0.05$ the \MD~phase remains only in the left corner of the parameter space, in the interval $\mu\lesssim 0.4\Lambda$ and $\sqrt{eB}\lesssim 0.25 \Lambda$. In both cases, the critical $eB$ decreases with increasing $\mu$.
All these effects are the manifestation of the IMRC effect induced by an interplay
between $\Omega R$ and $eB$. Apart from this aspect, a comparison between Figs. \ref{fig5}(d) and \ref{fig5}(e), the $\mu>m$ regime in the \MD~phase becomes also very narrow
(see the region between the solid red and dashed green lines in these figures), and a critical point appears in $(\mu,\sqrt{eB})\sim(0.03\Lambda,0.534\Lambda)$ for $\Omega R=0.01$ in Fig. \ref{fig5}(d), which is then shifted to  $(\mu,\sqrt{eB})\sim(0.069\Lambda,0.243\Lambda)$ in Fig. \ref{fig5}(e) for $\Omega R=0.05$. Black circles in Figs.\ref{fig5}(d) and \ref{fig5}(e) demonstrate the position of these critical points.	
\par
In Fig. \ref{fig6}, the phase portrait $\sqrt{eB}/\Lambda$-$\Omega R$ for fixed $\mu/\Lambda=0,0.2,0.4,0.6$ is plotted. In Fig.\ref{fig6}(a), for $\mu=0$ only the $m>\mu$ regime of the \MD~phase appears in the left corner of the parameter space. As it is shown, the critical value of $eB$ decreases with increasing $\Omega R$. This is an indication of the IMRC effect. The transition from the $\mu<m$ regime of the \MD~phase to the normal phase is of the first order. This confirms the results from Fig. \ref{fig2}, where, in particular, a first order phase transition occurs for $\Omega R\gtrsim 0.005$ in Figs. \ref{fig2}(c)-\ref{fig2}(e).
\\
By increasing $\mu$, an extremely narrow region with $\mu>m$ appears in the phase diagram. The red solid line in Fig. \ref{fig6}(b) separates two different regimes $\mu<m$ and $\mu>m$ in the \MD~phase, and the green dashed line separates the \MD~phase with $\mu>m$ from the normal phase.
In Fig. \ref{fig6}(c), by increasing $\mu$ to $\mu=0.4\Lambda$, the \MD~phase becomes further suppressed, so that it completely disappears for $\Omega R\gtrsim 0.06$. For $\mu=0.6\Lambda$, the $\mu<m$ regime of \MD~phase completely disappears from the parameter space, and its $\mu>m$ regime is shifted to the left corner, where the critical $eB$ does not significantly changes in terms of $\Omega R$ [see Fig. \ref{fig6}(d)]. This shows the interplay between $eB$, $\Omega R$ and $\mu$ in destroying the \MD~phase through the IMRC effect
\par
Finally, in Fig. \ref{fig7},  the phase portrait $\mu/\Lambda$-$\Omega R$ for fixed $\sqrt{eB}/\Lambda=0.15,0.25,0.45$ is plotted. The red solid lines separate two different regimes $\mu<m$ (denoted by "1") and $\mu>m$ (denoted by "2") in the \MD~phase, and the green dashed lines separate the \MD~phase with $\mu>m$ from the normal phase (denoted by "0"). As it is shown in Fig. \ref{fig7}(a), for small values of $\sqrt{eB}=0.25\Lambda$, the \MD~phase exists even for large values of $\Omega R\sim 0.1$.  The critical $\mu$ decreases with increasing $\Omega R$ [green dashed line in Fig. \ref{fig7}(a)]. For larger values of $\sqrt{eB}=0.45\Lambda$, the \MD~phase is completely shifted to the left corner of the parameter space. The $\mu>m$ regime in this phase is very narrow. A transition to the normal phase is only possible through this regime. The critical $\mu$ does not depend significantly on $\Omega R$. Let us remind that all green dashed lines correspond to the second-order phase transitions.
\section{Concluding remarks}\label{sec5}
\setcounter{equation}{0}
Starting with the Lagrangian density of a two-flavor gauged NJL model, and assuming a rigid  rotation about a certain axis with a constant angular velocity $\Omega$ and a magnetic field $eB$ aligned in the same direction, we defined two inhomogeneous condensates in order to introduce the \MD~phase in a rotating, dense, and magnetized cold quark matter.
The question was whether this phase survives the extreme conditions prevailing, e.g., in a neutron star. In contrast to \cite{incera2022}, the temperature and its possible interplay with high densities, constant magnetic fields and angular velocities does not play any role in the present paper. We consider a low temperature limit and showed that the temperature dependent thermodynamic potential vanishes in this limit. Taking this limit is justified by the assumption that the neutron star rotates with a constant angular velocity \cite{rezzolla-book}. 
\par
We first determined the energy eigenvalues of the model, and showed that in comparison with the nonrotating case, they are shifted by a term $\Omega j$, where $j=\ell+1/2$ is the corresponding quantum number to $J_{z}=L_{z}+\Sigma_{z}/2$ (see Sec. \ref{sec2} for more details). We then determined in Sec. \ref{sec3}, the corresponding thermodynamic potential to this model at finite temperature $T$, chemical potential $\mu$, and magnetic field $eB$. We showed that it consists of three parts, a vacuum part $V_{T=0}$, an $\mu,\Omega$ dependent part $V_{\mu,\Omega}$, and a temperature dependent part $V_{T\neq 0}$ [see \eqref{E13}]. The latter vanishes by taking the limit $T\to 0$. We thus focused on $V_{T=0}$ and $V_{\mu,\Omega}$.  As it is argued, $V_{T=0}$ is independent of $\Omega$. To regularize it, we used the same proper-time regularization scheme as in \cite{frolov2010,incera2022}. This gives us the possibility to compare our numerical results for $\Omega=0$ with the results presented in these two papers.
The final expression for $V_{T=0}$ is given in \eqref{E17}, where a summation over Landau levels is already performed. As concerns $V_{\mu,\Omega}$ from \eqref{E18}, we first separated it into two parts: The LLL and HLL parts, that are given in \eqref{E19} and \eqref{E20}. To derive the LLL part, we followed step by step the same regularization method as is described in the Appendix of \cite{frolov2010}. We showed, in particular, that the LLL part of the potential includes an anomalous term proportional to $b\Omega$, apart from a term proportional to $b\mu$, which appears originally in \cite{frolov2010} 
(see \eqref{E25}). This term
appears as a consequence of the asymmetry appearing in the energy dispersion relations of the lowest and higher Landau levels and, according to \cite{ferrer2022}, leads to important topological effects (see below for further discussions). 
Another difference comparing to the nonrotating case from \cite{frolov2010}, appears in the HLL part of the effective potential, whose final result which is given in \eqref{E33} depends explicitly on $\Omega$. 
\par
Minimizing the thermodynamic potential with respect to two dynamical variables $m$ and $b$, the chiral and spatial modulation condensates, we studied in Sec. \ref{sec4}, the $\mu, eB,$ and $\Omega R$ dependence of these two gaps, separately. We showed, that at $\mu=0$, nonvanishing $\Omega R$ acts as an additional chemical potential.\footnote{The fact that $\Omega N$, with $N=j_{\text{max}}$, is an ``effective'' chemical potential, is also indicated in \cite{fukushima2015}.}
For small $\Omega R$ up to $\Omega R\sim 0.005$, the rotation enhances the production of $b$  and thus the \MD~phase, even for vanishing $\mu$. Once $\Omega R$ and $eB$ become larger, $b$ vanishes because of the IMRC effect and the resulting suppression of $m$ (see Fig. \ref{fig2}). For nonvanishing $\mu$, the situation is worse.  Here, the IMRC effect is amplified because of the similar roles played by $\Omega R$ and $\mu$. They act as inverse magnetic catalyzers (magnetic inhibitors). According to the results from Fig. \ref{fig4}, we conclude that, because of the IMRC effect, the \MD~phase is suppressed in the regimes of $eB, \mu,$ and $\Omega R$ that are relevant for cold neutron stars.
\par
In the second part of Sec. \ref{sec4}, we further explored the $eB$-$\mu$, $eB$-$\Omega R$, and $\mu$-$\Omega R$ phase portraits.  We showed that, as expected from \cite{frolov2010}, the \MD~phase consists of two regimes: $\mu<m$ and $\mu>m$. The transition from the first regime $\mu<m$ to the normal (chiral symmetry restored) phase is of the first order, whereas the transition from the second regime $\mu>m$ to the normal phase is of the second-order.
The results from this part confirm the conclusion concerning the role of IMRC effect on destroying the \MD~phase in regimes that are relevant for the cold neutron stars. The fact that the critical magnetic field decreases with increasing $\Omega R$ ($\mu$) for fixed $\mu$ ($\Omega R$) is a direct consequence of the IMRC effect.
\par
Because of the aforementioned similarity between $\mu$ and $\Omega R$, the \MD~phase is created even at $\mu=0$ for $\Omega R>0$. This is in particular demonstrated in the plots of Fig. \ref{fig2} and the phase portrait in Fig. \ref{fig6}(a). According to these results, the critical value of $eB$ decreases with increasing $\Omega R$ for $\mu\leq 0.4\Lambda$. This opens the possibility for the \MD~phase to be created during the early stages of heavy ion collisions, where $\mu\sim 0$, $\sqrt{eB}\sim 0.1-0.5$ GeV ($eB\sim 5-15 m_{\pi}^{2}$ with the pion mass $m_{\pi}\sim 140$ MeV) \cite{huang2022}, and $\Omega R\sim 0.1$ \cite{becattini2016}. Here, however, it is necessary to consider the effect of finite temperature and its interplay with the magnetic field and rotation. First results in this direction is recently presented in \cite{tabatabaee2023}, where the effect of magnetic fields is neglected. In \cite{chernodub2020}, the effects of rotation on confining properties of compact electrodynamics in two spatial dimensions is studied. It is, in particular, shown that at finite temperature the phase diagram of a uniformly rotating system possesses, in addition to a confining phase at low temperature and a deconfining phase at high temperature, a mixed inhomogenous phase at intermediate temperatures. The latter has a confining region at the core and a deconfining region at the edge of the rotating system. It would be interesting to extend the present work in this direction.
\par
As aforementioned, in this paper, we focused on the \MD~phase in rotating cold quark matter. This phase is characterized by a certain asymmetry in its LLL spectrum. As it is shown in \cite{ferrer2015}, this asymmetry which occurs also in the nonrotating case, arises by the interplay between the inhomogeneous condensate and the magnetic field. It is also shown to be the origin of the nontrivial topological properties of this model, leading, in particular, to the creation of an anomalous, nondissipative electric Hall current in the corresponding Maxwell equations. As it is argued in \cite{ferrer2015}, in the nonrotating case, the anomalous baryon (quark) number density $\rho_{\text{an}}^{B}$ associated with this anomalous four-current is related to the regularized Atiyah-Parodi-Singer (APS) index $\eta_{B}$ \cite{semenoff1986} via
\begin{eqnarray*}
\eta_{B}\equiv \lim\limits_{s\to 0}\sum_{k}\mbox{sgn}\left(E_{k}\right)|E_{k}|^{-s}= -2\int d^{3}x \rho_{\text{an}}^{B}, \nonumber\\
\end{eqnarray*}
with
\begin{eqnarray}\label{A1}
\rho_{\text{an}}^{B}\equiv \sum_{f}\frac{N_{c}b|q_{f}eB|}{2\pi^{2}}.
\end{eqnarray}
Here, $b$ is the spatial modulation in the \MD~phase and $N_{c}$ the number of colors. According to \cite{ferrer2015}, another way to determine $\rho_{\text{an}}^{B}$  is by appropriately regularizing the anomalous part of the thermodynamic potential corresponding to the \MD~phase in a nonrotating phase \cite{frolov2010}. According to \cite{frolov2010},
\begin{eqnarray*}
\mathscr{V}_{\text{anomal}}\big|_{\Omega=0}=-\rho_{\text{an}}^{B}\mu,
\end{eqnarray*}
with baryonic charge density given in \eqref{A1}.
As concerns the rotating case, we determined in Sec. \ref{sec3}, the anomalous part of the thermodynamic potential of the \MD~model in a rotating medium. It is given in \eqref{E25} and leads to
\begin{eqnarray*}
\mathscr{V}_{\text{anomal}}\big|_{\Omega\neq 0}=-\left(\rho_{\text{an}}^{B}\mu+\rho_{\text{an}}^{\Omega} \Omega\right),
\end{eqnarray*}
where $\rho_{\text{an}}^{B}$ is given in \eqref{A1}, and $\rho_{\text{an}}^{\Omega}$ reads
\begin{eqnarray}\label{A2}
\rho_{\text{an}}^{\Omega}=\sum_{f}\mbox{sgn}(q_{f})\frac{N_{c}b|q_{f}eB|}{8\pi^{2}}\frac{R^{2}}{L_{f,B}^{2}},
\end{eqnarray}
with $R$ the radius of the rotating cylinder and the magnetic length $L_{f,B}\equiv|q_{f}eB|^{-1/2}$. Let us notice that $\rho_{\text{an}}^{\Omega}$ arises from the summation over a flavor dependent $\ell$ in \eqref{E23} [for the limits of $\ell$, see \eqref{E6}], where $\Omega j$ played the role of a chemical potential associated with rotation \cite{fukushima2015}. In analogy to $\rho_{\text{an}}^{B}$, $\rho_{\text{an}}^{\Omega}$ from \eqref{A2} leads immediately to anomalous electric charge density $J^{0}_{\Omega}\equiv\sum_{f}eq_{f}\rho_{\text{an},f}^{\Omega}$ associated to $\rho_{\text{an},f}^{\Omega}$ for each flavor, which may appear in the corresponding Maxwell equation as its baryonic counterpart $J^{0}_{B}\equiv\sum_{f}eq_{f}\rho_{\text{an},f}^{B}$ in \cite{ferrer2015}. It would be interesting to find the relation between $\rho_{\text{an}}^{\Omega}$ and the APS index $\eta$ via an appropriate regularization of $\eta\equiv \lim\limits_{s\to 0}\sum_{k}\mbox{sgn}\left(E_{k}\right)|E_{k}|^{-s}$ in the presence of $\Omega$. We postpone this computation to our future works.
\section*{Acknowledgements}
The authors thank M. H. Gholami for his collaboration in the early stages of this work. We also thank M. Sh. Sadeghi and Sh. Baghram for valuable discussions.
\begin{appendix}
\section{The spectrum of the model}\label{appA}
\setcounter{equation}{0}
In this appendix, we solve the eigenvalue equation $\mathscr{H}_{f}\psi_{f}=E_{f}\psi_{f}$ from \eqref{N11} in an appropriate cylindrical coordinate system, described by $x^{\mu}=(t,x,y,z)=(t,\rho\cos\varphi, \rho\sin\varphi,z)$, where $\rho$ is the radial coordinate, $\varphi$ the azimuthal angle, and $z$ the height.
\par
To do this, let us first consider $\mathscr{H}_{f}$ from \eqref{N12},
\begin{eqnarray}\label{appA1}
\mathscr{H}_{f}=\left(
\begin{array}{cc}
\mathscr{O}+bs_f\tau_{3}-\Omega J_{z}&m\\
m&-\mathscr{O}+bs_f\tau_{3}-\Omega J_{z}
\end{array}
\right),\nonumber\\
\end{eqnarray}
with $\mathscr{O}$ given by
\begin{eqnarray}\label{appA2}
\mathscr{O}\equiv \tau_1\left(i\partial_{x}-q_{f}eBy/2\right)+\tau_2\left(i\partial_{y}+q_{f}eBx/2\right)+i\tau_{3}\partial_{z}.\nonumber\\
\end{eqnarray}
Using the representation of the total angular momentum in the cylindrical coordinate system $J_{z}=L_{z}+\Sigma_{z}/2=-i\partial_{\varphi}+\Sigma_{z}/2$, with $\Sigma_{z}=\mathbb{I}_{2\times 2}\otimes \tau_{3}$, it is easy to check that $\mathscr{H}_{f}$ commutes with $J_{z}$.They thus have simultaneous eigenfunctions. The latter has the general form
\begin{eqnarray}\label{appA3}
\psi_{f}\left(\rho,\varphi,z\right)= e^{ip_z z}\left(
\begin{array}{c}
e^{i\ell_{-}\varphi}f_1(\rho) \\
e^{i\ell_{+}\varphi}f_2(\rho) \\
e^{i\ell_{-}\varphi}f_3(\rho) \\
e^{i\ell_{+}\phi}f_4(\rho)
\end{array}
\right),
\end{eqnarray}
and satisfies the eigenvalue equation for the total angular momentum,
\begin{eqnarray}\label{appA4}
J_{z}\psi_{f}\left(\rho,\varphi,z\right)=j\psi_{f}\left(\rho,\varphi,z\right).
\end{eqnarray}
Here, $\ell_{\pm}\equiv j\pm 1/2$.
To determine $f_{i}(\rho), i=1,\cdots 4$, let us consider the energy eigenfunction equation  \eqref{N11}, with $\mathscr{H}_{f}$ given in \eqref{appA1}. Plugging $\psi_{f}\left(\rho,\varphi,z\right)$ from \eqref{appA3} into this relation
we arrive first at
\begin{eqnarray}\label{appA5}
\left(\mathscr{H}_{f}+\Omega J_z\right)\psi_{f}=\mathscr{M}_{f}\psi_{f},
\end{eqnarray}
with $\mathscr{M}_{f}$ given by
\begin{eqnarray}\label{appA6}
\mathscr{M}_{f}\equiv\left(
\begin{array}{cc}
\mathscr{D}_f^{+}&m\\
m&\mathscr{D}_f^{-}
\end{array}
\right),
\end{eqnarray}
and
\begin{widetext}
\begin{eqnarray}\label{appA7}
\mathscr{D}_{f}^{\pm} =
		\begin{pmatrix}
			\mp p_z+bs_f & \pm ie^{-i\varphi}(\partial_\rho - \frac{i}{\rho}\partial_\varphi)\mp i\frac{q_{f} eB}{2}\rho e^{-i\phi}\\
			\pm ie^{i\varphi}(\partial_\rho +\frac{i}{\rho}\partial_\varphi) \pm  i\frac{q_{f}eB}{2}\rho e^{i\varphi} & \pm p_z-bs_f
		\end{pmatrix}.
\end{eqnarray}
\end{widetext}
From \eqref{appA5}, we obtain
\begin{eqnarray}\label{appA8}
[\left(\mathscr{H}_{f}+\Omega J_z\right)^{2}-m^{2}]\psi_{f}=\mathscr{N}_{f}\psi_{f},
\end{eqnarray}
with
\begin{eqnarray}\label{appA9}
\mathscr{N}_{f}\equiv
\begin{pmatrix}
		\mathscr{E}_{f}^{--} & 0 & 2mbs_f  & 0    \\
		0 & \mathscr{E}_{f}^{+-} & 0 & -2mbs_f \\
		2mbs_f & 0 & \mathscr{E}_{f}^{-+}  & 0 \\
		0 & -2mb s_f & 0 & \mathscr{E}_{f}^{++}
	\end{pmatrix},
\end{eqnarray}
and
\begin{eqnarray}\label{appA10}
	\mathscr{E}_{f}^{\mp \mp}&\equiv&   -\partial_{\rho}^{2} -\frac{1}{\rho}\partial_{\rho}-\frac{1}{\rho^2}\partial^2_\phi +iq_{f}eB\partial_\phi \mp q_{f} eB \nonumber\\
&&+\left(\frac{q_{f} eB\rho}{2}\right)^2 +(p_z \mp s_fb)^2,\nonumber\\
	\mathscr{E}_{f}^{\pm \mp}&\equiv&   -\partial_{\rho}^{2} -\frac{1}{\rho}\partial_{\rho}-\frac{1}{\rho^2}\partial^2_\phi +iq_{f}eB\partial_\phi \pm q_{f} eB \nonumber\\
&&+\left(\frac{q_{f} eB\rho}{2}\right)^2 +(p_z \mp s_fb)^2.
\end{eqnarray}
As it turns out, the above matrix $\mathscr{N}_{f}$ commutes with the diagonal matrix
\begin{eqnarray}\label{appA11}
\widetilde{\mathscr{N}}_{f}\equiv
\begin{pmatrix}
		\widetilde{\mathscr{E}}_{f}^{-} & 0 & 0 & 0    \\
		0 &\widetilde{\mathscr{E}}_{f}^{+} & 0 & 0 \\
		0& 0 & \widetilde{\mathscr{E}}_{f}^{-}  & 0 \\
		0 & 0& 0 &\widetilde{\mathscr{E}}_{f}^{+}
	\end{pmatrix},
\end{eqnarray}
with
\begin{eqnarray}\label{appA12}
		\widetilde{\mathscr{E}}_{f}^{\mp}&=& -\partial_\rho^2 -\frac{1}{\rho}\partial_\rho-\frac{1}{\rho^2}\partial^2_\varphi +iq_{f}eB\partial_\varphi \mp q_{f}eB\nonumber\\
&&+\left(\frac{q_f eB\rho}{2}\right)^{2}.
\end{eqnarray}
This help us to determine the eigenfunction $\psi_{f}$ in \eqref{appA8}. Hence, the problem is reduced to determining the eigenfunction $\psi_{f}$ and the eigenvalue $\lambda_{f}$ in
\begin{eqnarray}\label{appA13}
\widetilde{\mathscr{N}}_{f}\psi_{f}=\lambda_{f}\psi_{f}.
\end{eqnarray}
Applying $\widetilde{\mathscr{N}}_{f}$ from \eqref{appA11} to $\psi_{f}$ from \eqref{appA3}, we arrive at two differential equations for $f_{i}(\rho), i=1,\cdots, 4$,
\begin{eqnarray}\label{appA14}
\lefteqn{\hspace{-3cm}\bigg[\partial_\rho^2 +\frac{1}{\rho}\partial_\rho-\frac{\ell_{\mp}^2}{\rho^2}  +q_{f}eB\left(\ell_{\mp}\pm 1\right) -\left(\frac{q_{f}eB\rho}{2}\right)^2} \nonumber\\
&&+\lambda_{f}\bigg]f_{\mp}(\rho)= 0,\nonumber\\
\end{eqnarray}
where $f_{-}\equiv f_{1}=f_{3}$ and $f_{+}\equiv f_{2}=f_{4}$ are introduced. To solve these equations, we use the ansatz
\begin{eqnarray}\label{appA15}
		f_{\mp}(x) = e^{-\frac{x}{2}} x^{|\ell_{\mp}|/2} g_{\mp}(x),
\end{eqnarray}
with $x\equiv |q_{f}eB|\rho^{2}/2$. Plugging \eqref{appA15} into \eqref{appA14}, the resulting equation reads
\begin{eqnarray}\label{appA16}
\big[x\partial_{x}^{2}+\left(|\ell_{\mp}|+1-x\right)\partial_{x}+\kappa_{\mp}\big]g_{\mp}(x)=0,
\end{eqnarray}
where
\begin{eqnarray}\label{appA17}
\kappa_{\mp}\equiv \frac{\lambda_{f}}{2|q_{f}eB|}+\frac{s_{f}\left(\ell_{\mp}\pm 1\right)-|\ell_{\mp}|-1}{2}.
\end{eqnarray}
The differential equation \eqref{appA16} is comparable with the Kummer's differential equation
\begin{eqnarray}\label{appA18}
\left(z\partial_{z}^{2}+(b-z)\partial_{z}-a\right)g(z)=0,
\end{eqnarray}
whose solution
\begin{eqnarray}\label{appA19}
g(z)=A{}_{1}F_{1}(a;b;z)+BU\left(a;b;z\right),
\end{eqnarray}
is a linear combination of a hypergeometric function of the first and second kind ${}_{1}F_{1}(a;b;z)$ and $U\left(a;b;z\right)$.\footnote{{Similar results are also found in \cite{fukushima2015, chernodub2017,sadooghi2021}.}} Requiring that $g_{\mp}(x)$ are regular at $x\to 0$, $g_{\mp}(x)$ is given by
\begin{eqnarray}\label{appA20}
g_{\mp}(x)=A_{\mp}{}_{1}F_{1}(-\kappa_{\mp};|\ell_{\mp}|+1,x).
\end{eqnarray}
Plugging this result into \eqref{appA15}, we thus arrive at
\begin{eqnarray}\label{appA21}
\hspace{-0.5cm}f_{\mp}(x) = A_{\mp}e^{-\frac{x}{2}} x^{|\ell_{\mp}|/2}{}_{1}F_{1}(-\kappa_{\mp};|\ell_{\mp}|+1,x).
\end{eqnarray}
Assuming, at this stage, that the fermionic system has no spatial boundary condition, the hypergeometric function can be replaced by the associated Laguerre polynomial,
\begin{eqnarray}\label{appA22}
\hspace{-0.5cm}{}_{1}F_{1}(-\kappa_{\mp};|\ell_{\mp}|+1,x)=\frac{|\ell_{\mp}|!\kappa_{\mp}!}{\left(|\ell_{\mp}|+\kappa_{\mp}\right)!}L_{\kappa_{\mp}}^{|\ell_{\mp}|}(x).
\end{eqnarray}
Moreover, introducing $n\equiv \frac{\lambda_{f}}{2|q_f eB|}$  in \eqref{appA17}, it turns out that in this case $n\in \mathbb{N}_{0}$. Using the orthonormality relation of the Laguerre polynomial
\begin{eqnarray}\label{appA23}
\int_{0}^{\infty}dz z^{\alpha}L_{n}^{\alpha}(z)L_{m}^{\alpha}(z)=\frac{(n+\alpha)!}{n!}\delta_{m,n},
\end{eqnarray}
we arrive at
\begin{eqnarray}\label{appA24}
f_{\mp}(x) = \left(\frac{|q_f eB|}{2\pi}\frac{\kappa_{\mp}}{\left(\kappa_{\mp}+|\ell_{\mp}|\right)!}\right)^{1/2}e^{-\frac{x}{2}} x^{|\ell_{\mp}|/2}L_{\kappa_{\mp}}^{|\ell_{\mp}|}(x).\nonumber\\
\end{eqnarray}
Plugging this result into \eqref{appA3}, the eigenfunction $\psi_{f}$ in \eqref{N11} reads
\begin{eqnarray}\label{appA25}
\psi_{f}(\rho,\varphi,z)=e^{ip_{z}z}\left(
\begin{array}{c}
e^{i\ell_{-}\varphi}f_{-}(x) \\
e^{i\ell_{+}\varphi}f_{+}(x) \\
e^{i\ell_{-}\varphi}f_{-}(x) \\
e^{i\ell_{+}\phi}f_{+}(x)
\end{array}
\right),
\end{eqnarray}
with $x=|q_{f}eB|\rho^2/2$ and $f_{\mp}$ from \eqref{appA24}. At this stage, we shall determine $\kappa_{\mp}$. To do this, we use the fact that the lower index $\kappa_{\pm}$ in the Laguerre polynomial shall be positive. According to \eqref{appA17}, depending on $\ell$, this leads for the two cases of positive and negative $q_{f}$, or alternatively positive and negative $s_{f}$, to different allowed values for $\kappa_{\mp}$. The results for $\ell\leq -1$ and $\ell\geq 0$ are summarized in Tables \ref{tab1} and \ref{tab2} (see also \cite{sadooghi2021} for more details).
\begin{center}
\begin{table}[ht]
    \centering
\caption{The allowed values for $\kappa_{s}, s\equiv\pm$ for $\ell\leq -1$, $\ell\geq 0$, and  positive and negative charges, according to \eqref{appA17}. }\label{tab1}    \vspace{1ex}
    \vspace{1ex}
        \begin{tabular}{c|rclcrcl}
            \hline\hline
&$\ell$&$\leq$&-1&\qquad\qquad&$\ell$&$\geq$&$0$\\
\hline
  $s_{f}=+1, s=-1$&$\kappa_{-}$&$=$&$n+\ell$&\qquad\qquad&$\kappa_{-}$&=&$n$ \\
  $s_{f}=+1, s=+1$&$\kappa_{+}$&$=$&$n+\ell$&\qquad\qquad&$\kappa_{+}$&=&$n-1$ \\
  $s_{f}=-1, s=-1$&$\kappa_{-}$&$=$&$n-1$&\qquad\qquad&$\kappa_{-}$&=&$n-\ell-1$ \\
  $s_{f}=-1, s=+1$&$\kappa_{+}$&$=$&$n$&\qquad\qquad&$\kappa_{+}$&=&$n-\ell-1$ \\
            \hline\hline
        \end{tabular}
\end{table}
\end{center}
\begin{table*}[ht]
    \centering
    \caption{The allowed values for $n=\frac{\lambda_{f}}{2|q_f eB|}$ for $s=\mp 1$, corresponding to $\kappa_{\mp}$, and  positive and negative charges, according to Table \ref{tab1}. To determine these ranges, we set $\kappa_{\mp}\geq 0$ in Table \ref{tab1}.}
    \label{tab2}
    \vspace{1ex}
        \begin{tabular}{c|rclcl}
            \hline\hline
            \multirow{2}{*}{$s_{f}=+1, s=-1$}
            &$n$&$=$&$0$&\qquad\qquad&$\ell=0,1,2,\cdots$\\
            &$n$&$\geq$&$1$&\qquad\qquad&$\ell=-n,-n+1,\cdots,-2,-1,0,1,2,\cdots$ \\
            \hline
            \multirow{2}{*}{$s_{f}=+1,s=+1$}
            &$n$&$=$&$0$&\qquad\qquad&---\\
            &$n$&$\geq$&$1$&\qquad\qquad&$\ell=-n,-n+1,\cdots,-2,-1,0,1,2,\cdots$ \\
            \hline\hline
            \multirow{2}{*}{$s_{f}=-1,s=-1$}
            &$n$&$=$&$0$&\qquad\qquad&---\\
            &$n$&$\geq$&$1$&\qquad\qquad&$\ell=\cdots,-2,-1,0,1,2,\cdots,n-2,n-1$ \\
            \hline
            \multirow{2}{*}{$s_{f}=-1,s=+1$}
            &$n$&$=$&$0$&\qquad\qquad&$\ell=\cdots,-2,-1$\\
            &$n$&$\geq$&$1$&\qquad\qquad&$\ell=\cdots,-2,-1,0,1,2,\cdots,n-2,n-1$ \\
            \hline\hline
        \end{tabular}
\end{table*}
According to the results from Table \ref{tab2}, the energy eigenfunctions for $n=0$ are different from the ones for $n\neq 0$: For $n=0$ and $s_{f}=+1$ (positive charges), we have
\begin{eqnarray}\label{appA26}
\psi_{f}^{n=0,s_{f}=+1}=e^{ip_{z}z}e^{i\ell_{-}\varphi}f_{-}(x)\left(
\begin{array}{c}
1 \\
0\\
1 \\
0
\end{array}
\right),
\end{eqnarray}
whereas for $s_{f}=-1$ (negative charges), we obtain
\begin{eqnarray}\label{appA27}
\psi_{f}^{n=0,s_{f}=-1}=e^{ip_{z}z}e^{i\ell_{+}\varphi}f_{+}(x)\left(
\begin{array}{c}
0 \\
1\\
0 \\
1
\end{array}
\right).
\end{eqnarray}
For $n\geq 1$, the eigenfunctions are given by \eqref{appA25}.
\par
To determine the energy eigenvalues $E_{f}$, we have to distinguish between $n=0$ and $n\neq 0$ for positive ($s_{f}=+1$) and negative ($s_{f}=-1$) charges.  For $n=0$ and $s_{f}=+1$ as well as $s_f=-1$, we plug \eqref{appA26} as well as \eqref{appA27} into \eqref{appA5} and use \eqref{N11} to arrive at
\begin{eqnarray}\label{appA28}
\lefteqn{\hspace{-1cm}
\begin{pmatrix}
\mp p_z +b & m\\
m & \pm p_z+b
\end{pmatrix}
\psi_{f}^{n=0,s_{f}=\pm 1}
}\nonumber\\
&&
= (E_f+\Omega j) \psi_{f}^{n=0,s_{f}=\pm 1}.
\end{eqnarray}
Here, we have reduced the $4\times 4$ matrix equation to a $2\times 2$ one. This leads to
the energy dispersion relation
\begin{eqnarray}\label{appA29}
E_{f}^{n=0}=-\Omega j+b+\epsilon\sqrt{p_{z}^{2}+m^{2}},
\end{eqnarray}
for $n=0$ (see \eqref{N13}). For $n>0$, we plug \eqref{appA25} into \eqref{appA8} and combine the result with \eqref{appA13} and \eqref{N11} to arrive first at
\begin{eqnarray}\label{appA30}
\begin{pmatrix}
\mathscr{K}_{-s_{f}}&0&+2ms_{f}b&0\\
0&\mathscr{K}_{+s_{f}}&0&-2ms_{f}b\\
+2ms_{f}b&0&\mathscr{K}_{-s_{f}}&0\\
0&-2ms_{f}b&0&\mathscr{K}_{+s_{f}}
\end{pmatrix}\psi_{f}=0,\nonumber\\
\end{eqnarray}
with
\begin{eqnarray}\label{appA31}
\mathscr{K}_{\mp}\equiv 2n|q_{f}eB|+(p_{z}\mp s_{f}b)^{2}-[(E_{f}+\Omega j)^{2}-m^2]. \nonumber\\
\end{eqnarray}
This leads to the energy dispersion relation
\begin{eqnarray}\label{appA32}
E_{f}^{n>0}=-\Omega j+\zeta\bigg[\left(b+\epsilon\sqrt{p_{z}^{2}+m^2}\right)^{2}+2n|q_{f}eB|\bigg]^{1/2},\nonumber\\
\end{eqnarray}
for $n>0$ (see \eqref{N14}).
\section{Supplementary results}\label{appB}
\setcounter{equation}{0}
\subsection{The $p_{z}$ integration in \eqref{E21}}\label{appB1}
Let us consider the integration over $p_{z}$, which appears in $\mathscr{V}_{\mu,\Omega}^{n=0}$ from \eqref{E21},
\begin{eqnarray}\label{appB1}
\mathscr{V}_{f,\ell}=\sum_{\epsilon=\pm 1}\int_{-\infty}^{\infty}dp_{z}\left(|E_{f}^{n=0}-\mu|-|E_{f}^{n=0,\Omega=0}|\right).\nonumber\\
\end{eqnarray}
Using the definition of $E_{f}^{n=0}$, and performing the summation over $\epsilon$, we first arrive at
\begin{eqnarray}\label{appB2}
\mathscr{V}_{f,\ell}=2 \left(J^{\text{mom}}
\left(a_{1}\right)-J^{\text{mom}}\left(a_{2}\right)\right),
\end{eqnarray}
with
\begin{eqnarray}\label{appB3}
J^{\text{mom}}(a)\equiv \int_{0}^{\Lambda}dp_{z}\left(\big|a+\omega(p_{z})\big|+\big|a-\omega(p_{z})\big|\right),\nonumber\\
\end{eqnarray}
and $a_{1}\equiv b-\Omega j-\mu$ and $a_{2}\equiv b$. In \eqref{appB3}, $\omega(p)$ is defined by $\omega(p)\equiv \sqrt{p^{2}+m^{2}}$. Using
\begin{widetext}
\begin{eqnarray}\label{appB4}
\int_{0}^{\Lambda}dp_{z}\big|a+\omega(p_{z})\big|&=&\left\{
\begin{array}{lcrcl}
I\left(0,\Lambda\right)+a\Lambda,&\qquad&a&>&-|m|,\\
-I(0,P_{a})+I\left(P_{a},\Lambda\right)-2aP_{a}+a\Lambda,&\qquad&a&<&-|m|,\\
\end{array}
\right.\nonumber\\
\int_{0}^{\Lambda}dp_{z}\big|a-\omega(p_{z})\big|&=&\left\{
\begin{array}{lcrcl}
-I(0,P_{a})+I\left(P_{a},\Lambda\right)+2aP_{a}-a\Lambda,&\qquad&a&>&|m|,\\
I\left(0,\Lambda\right)-a\Lambda,&\qquad&a&<&|m|,\\
\end{array}
\right.
\end{eqnarray}
with $P_{a}\equiv \sqrt{a^{2}-m^{2}}$, and
\begin{eqnarray}\label{appB5}
I(\Lambda_{1},\Lambda_{2})\equiv\int_{\Lambda_{1}}^{\Lambda_{2}}dp_{z}\omega(p_{z})=\frac{1}{2}\bigg[\Lambda_{2}\omega(\Lambda_{2})-\Lambda_{1}\omega(\Lambda_{1})+m^{2}\ln\left(\frac{\Lambda_{2}+\omega(\Lambda_{2})}{\Lambda_{1}+\omega(\Lambda_{1})}\right)\bigg],
\end{eqnarray}
we arrive at \cite{frolov2010}
\begin{eqnarray}\label{appB6}
J^{\text{mom}}(a)=\left\{
\begin{array}{lcc}
2I(P_{a},\Lambda)+2aP_{a},&\qquad&|m|<a,\\
2I(0,\Lambda),&\qquad&-|m|<a<|m|,\\
2I(P_{a},\Lambda)-2aP_{a},&\qquad&a<-|m|.
\end{array}
\right.
\end{eqnarray}
Plugging \eqref{appB6} into \eqref{appB2}, we obtain \eqref{E22}.
\subsection{The final expression for $\mathscr{K}_{f,\epsilon,n}(p_z)$ in \eqref{E33}}\label{appB2}
In this section, we present the final expression for $\mathscr{K}_{f,\epsilon,n}(p_{z})$ defined by
\begin{eqnarray}\label{appB7}
\mathscr{K}_{f,\epsilon,n}(p_{z})\equiv \sum_{k=0}^{N_{f}-1}\mathscr{J}_{f,\epsilon,n,k}(p_{z}),
\end{eqnarray}
where $\mathscr{J}_{f,\epsilon,n,k}(p_{z})$ is defined in \eqref{E31}. We perform the summation over $k$ in \eqref{appB7} using an appropriate Mathematica program and arrive at the following conditional expression:
\begin{eqnarray}\label{appB8}
\mathscr{K}_{f,\epsilon,n}(p_{z})=
\mathcal{A}_{i}\qquad\mbox{if}\quad\mathcal{C}_{i},\quad \mbox{for}\quad i=1,\cdots, 10,
\end{eqnarray}
with
\begin{eqnarray}\label{appB9}
\mathcal{A}_{1}&\equiv& -\Omega E_{+}^{+},\nonumber\\
\mathcal{A}_{2}&\equiv&-\frac{1}{2}\Omega\mathcal{N}_{f}(2 E_{+}^{-}-(\mathcal{N}_{f}-1)),\nonumber\\
\mathcal{A}_{3}&\equiv&-\frac{1}{2}\Omega\mathcal{N}_{f}(2 E_{+}^{+}+(\mathcal{N}_{f}-1)),\nonumber\\
\mathcal{A}_{4}&\equiv&-\frac{1}{2}\Omega\left(\left\lceil E_{+}^{-}\right\rceil-\mathcal{N}_{f} \right)\left(\left\lceil E_{+}^{-}\right\rceil -\left(2E_{+}^{-}+1-\mathcal{N}_{f}\right)\right),\nonumber\\
\mathcal{A}_{5}&\equiv&-\frac{1}{2}\Omega \left(\left(\left\lceil E_{+}^{-}\right\rceil-(2E_{+}^{-}+1)\right) \left\lceil E_{+}^{-}\right\rceil +
2E_{+}^{+}+\mathcal{N}_{f}\left(2E_{+}^{-}+1-\mathcal{N}_{f}\right)\right),\nonumber\\
 \mathcal{A}_{6}&\equiv&-\frac{1}{2}\Omega \left(2 E_{+}^{+}-\left\lceil E_{+}^{+}\right\rceil \right) \left(\left\lfloor -E_{+}^{+}\right\rfloor +1\right),\nonumber\\
\mathcal{A}_{7}&\equiv&-\frac{1}{2}\Omega\left(\left( \left\lceil E_{+}^{-}\right\rceil-(2 E_{+}^{-}+1)\right) \left\lceil E_{+}^{-}\right\rceil+  \left(\left\lceil E_{+}^{+}\right\rceil -(2 E_{+}^{+}+1)\right) \left\lceil E_{+}^{+}\right\rceil
+2\left(E_{+}^{+}+E_{+}^{-}\mathcal{N}_{f}\right)+\mathcal{N}_{f}\left(1-\mathcal{N}_{f}\right)
 \right),\nonumber\\
\mathcal{A}_{8}&\equiv&-\frac{1}{2}\Omega\left(\left\lceil E_{+}^{-}\right\rceil -\left\lfloor E_{+}^{-}\right\rfloor -1\right)\left(\left\lceil E_{+}^{-}\right\rceil +\left\lfloor E_{+}^{-}\right\rfloor -2E_{+}^{-}\right),\nonumber\\
\mathcal{A}_{9}&\equiv&-\frac{1}{2}\Omega \left(\left(\left\lceil E_{+}^{-}\right\rceil-(2 E_{+}^{-}+1)\right) \left\lceil E_{+}^{-}\right\rceil -\left(\left\lfloor E_{+}^{-}\right\rfloor-(2E_{+}^{-}-1)\right) \left\lfloor E_{+}^{-}\right\rfloor +4E_{+}^{0}\right),\nonumber\\
\mathcal{A}_{10}&\equiv&-\frac{1}{2}\Omega \left(\left(\left\lceil E_{+}^{-}\right\rceil -(2 E_{+}^{-}+1)\right) \left\lceil E_{+}^{-}\right\rceil +\left(\left\lceil E_{+}^{+}\right\rceil -(2E_{+}^{+}+1) \right)\left\lceil E_{+}^{+}\right\rceil -\left(\left\lfloor E_{+}^{-}\right\rfloor-\left(2E_{+}^{-}-1\right)\right)\left\lfloor E_{+}^{-}\right\rfloor +4 E_{+}^{0}\right),\nonumber\\
\end{eqnarray}
and the conditions
\begin{eqnarray}\label{appB10}
\mathcal{C}_{1}&\equiv&E_{+}^{+}=0\land \mu_{f}<0\land E_{+}^{-}>\mathcal{N}_{f}-1
\land E_{+}>0\land \mathcal{N}_{f}>1\land \Omega >0,\nonumber\\
\mathcal{C}_{2}&\equiv&E_{+}^{-}<0\land E_{+}>0\land \mathcal{N}_{f}>1\land E_{+}^{+}>0\land \Omega >0,\nonumber\\
\mathcal{C}_{3}&\equiv&\mu_{f}<0\land E_{+}^{+}<0\land
E_{+}^{-}>\mathcal{N}_{f}-1
\land E_{+}>0\land \mathcal{N}_{f}>1\land \Omega >0\land E_{+}^{+}\leq 1-\mathcal{N}_{f},\nonumber\\
\mathcal{C}_{4}&\equiv&E_{+}>0\land \mathcal{N}_{f}>1\land E_{+}^{-}\geq 0\land
E_{+}^{-}<\mathcal{N}_{f}-1
\land \big[E_{+}^{+}>0\lor (E_{+}^{+}=0\land \Omega \leq 0)\big],\nonumber\\
\mathcal{C}_{5}&\equiv&E_{+}>0\land \mu_{f}<0\land \mathcal{N}_{f}>1\land \Omega >0\land E_{+}^{+}=0\land
E_{+}^{-}<\mathcal{N}_{f}-1,\nonumber\\
\mathcal{C}_{6}&\equiv&\mu_{f}<0\land E_{+}^{+}<0\land
E_{+}^{+}>1-\mathcal{N}_{f}
\land E_{+}^{-}>\mathcal{N}_{f}-1
\land E_{+}>0\land \mathcal{N}_{f}>1,\nonumber\\
\mathcal{C}_{7}&\equiv&E_{+}>0\land \mu_{f}<0\land \mathcal{N}_{f}>1\land E_{+}^{+}<0\land
E_{+}^{-}<\mathcal{N}_{f}-1,\nonumber\\
\mathcal{C}_{8}&\equiv&E_{+}^{-}=\mathcal{N}_{f}-1\land E_{+}>0\land \mathcal{N}_{f}>1\land E_{+}>\mu_{f} \nonumber\\
&& \land\left(\left(\Omega \leq 0\land\left(E_{+}+\mu_{f}\geq 0\lor \Omega [E_{+}^{+}+(\mathcal{N}_{f}-1)]\leq 0\right)\right)\lor E_{+}^{+}>0\right),\nonumber\\
\mathcal{C}_{9}&\equiv&E_{+}>0\land \mu_{f}<0\land \mathcal{N}_{f}>1\land \Omega >0\land E_{+}^{+}=0\land E_{+}^{-}=\mathcal{N}_{f}-1,\nonumber\\
\mathcal{C}_{10}&\equiv&\mathcal{N}_{f}>1\land E_{+}>0\land \mu_{f}<0\land
E_{+}^{-}=\mathcal{N}_{f}-1
\land E_{+}^{+}<0\land E_{+}^{+}>1-\mathcal{N}_{f}.
\end{eqnarray}
In the above expressions, $E^{\pm,0}$ are defined by
\begin{eqnarray}
E_{+}^{\pm}\equiv\frac{E_{+}\pm\mu_{f}}{\Omega},\qquad
E_{+}^{0}\equiv\frac{E_{+}}{\Omega}.
\end{eqnarray}
Here, $E_{+}$ and $\mu_{f}$ for $f=\{u,d\}$ are defined in \eqref{E27} and \eqref{E32}.
\end{widetext}
\end{appendix}

\end{document}